\documentclass
[english,manuscript,prd,unsortedaddress,showpacs,titlepage,nofootinbib]{revtex4}%
\usepackage{amsmath}
\usepackage[T1]{fontenc}
\usepackage[latin9]{inputenc}
\usepackage{amssymb}
\usepackage{babel}
\usepackage{appendix}
\usepackage{lscape}
\usepackage{graphicx}
\usepackage{amsfonts}%
\providecommand{\U}[1]{\protect\rule{.1in}{.1in}}
\begin{document}
\title[near-circular, near-polar orbits]{The Carter Constant for Inclined Orbits About a Massive Kerr Black Hole:
near-circular, near-polar orbits }
\author{P. G. Komorowski}
\email{pkomoro@uwo.ca}
\affiliation{Department of Physics and Astronomy,\\University of Western Ontario,\\1151 Richmond Street,\\London, Ontario, Canada, N6A 3K7}
\author{S. R. Valluri}
\email{valluri@uwo.ca}
\affiliation{Department of Physics and Astronomy }
\affiliation{Department of Applied Mathematics,\\University of Western Ontario,\\1151 Richmond Street,\\London, Ontario, Canada, N6A 3K7}
\author{M. Houde}
\email{mhoude2@uwo.ca}
\affiliation{Department of Physics and Astronomy,\\University of Western Ontario,\\1151 Richmond Street,\\London, Ontario, Canada, N6A 3K7}

\begin{abstract}
In an extreme mass-ratio binary black hole system, a non-equatorial orbit will
list (i.e. increase its angle of inclination, ${\iota}$) as it evolves in Kerr
spacetime. \ The abutment, a set of evolving, near-polar, retrograde orbits,
for which the instantaneous Carter constant (${Q)}$ is at its maximum value
(${Q}_{X}$) for given values of latus rectum (${\tilde{l}}$) and eccentricity
(${e}$), has been introduced as a laboratory in which the consistency of
$dQ/dt$ with corresponding evolution equations for $d\tilde{l}/dt$ and $de/dt$
might be tested independently of a specific radiation back-reaction model. To
demonstrate the use of the abutment as such a laboratory, a derivation of
$dQ/dt$, based only on published formulae for $d\tilde{l}/dt$ and $de/dt$, was
performed for elliptical orbits on the abutment. \ The resulting expression
for $dQ/dt$ matched the published result to the second order in $e$. \ We
believe the abutment is a potentially useful tool for improving the accuracy
of evolution equations to higher orders of $e$ and $\tilde{l}^{-1}$.

\end{abstract}

\pacs{04.70.Bw,04.30.Db}
\maketitle

\section{Introduction}

An extreme mass-ratio binary black hole system (EMRI) is composed of a primary
object, which can be a Kerr black hole of mass $M\sim10^{6}-10^{7}$ solar
masses with a spin \footnote{Our use of ${\tilde{S}}$ for black hole spin
arose during our initial studies of the work of Barack and Cutler
\cite{Barack:2004uq}.} ${\tilde{S}}=\left\vert {{\mathbf{J}}}\right\vert
/{M}^{2}$ (where ${{\mathbf{J}}}$ is the spin angular momentum), and an
orbiting secondary object of mass $m\sim1-10$ solar masses. Theoretical models
to describe the orbital evolution of the secondary object in various
situations have been derived and presented in the literature: circular orbits
in the equatorial plane of the primary object
\cite
{1973blho.conf..215B,1993PhRvD..47.1497P,1993PhRvD..47.1511C,Ryan:1995ix,2000PhRvD..62l4022O,2011arXiv1105.2959P}%
, elliptical orbits in the equatorial plane
\cite
{PhysRev.131.435,PhysRev.136.B1224,1991STIA...9240399B,1992MNRAS.254..146J,PhysRevD.50.3816,2002PhRvD..66d4002G,2005PhRvL..94v1101H}%
, and an extensive body of research on circular or elliptical orbits inclined
with respect to the equatorial plane
\cite
{1973ApJ...185..635T,1973ApJ...185..649P,1995PhLA..202..347O,Ryan:1996ly,1996PhRvD..53.4319K,PhysRevD.55.3444,1999CQGra..16.2929D,PhysRevD.61.084004,2001PhRvD..64f4004H,2002PhRvD..66f4005G,Barack:2004uq,2005PThPh.114..509S,2006PThPh.115..873S,2006PhRvD..73f4037G,2007PhRvD..76d4007B,2007PThPh.117.1041G,Flanigan:2007kx,2007PhRvD..75b4005B}%
. Such models are used to generate hypothetical gravitational waveforms (GW),
which provide templates for use in the detection of gravitation wave signals
by pattern recognition (Punturo et al. \cite{Punturo:2010zza}). The detection
of GW radiation by the Earth-based Laser Interferometer Gravitational Wave
Observatory\ (LIGO) or the Laser Interferometer Space Antenna (LISA) depends
fundamentally on the availability of correct templates
\cite{2005PhRvL..94v1101H,Cutler:2007mi,Normandin:2008:GWS:1460936.1461188}.

Performing direct observations of relativistic effects is an important
challenge. The Solar System affords one the opportunity to observe and model
the motions of natural and artificial bodies in Kerr spacetime in the
weak-field, slow motion limit \cite{Iorio2011AstroSpaceSci,EverittPRL2011};
and recent measurements of artificial-satellite orbits have produced estimates
of the Lense-Thirring precession to an accuracy of 10\%
\cite{Iorio2011AstroSpaceSci}. Further, the discovery of Sagittarius A*, a
massive black hole (MBH) of $\sim4.0\times10^{6}$ solar masses, at the centre
of our galaxy (see \cite{IorioPRD2011,SadeghianCQG2011,IorioMNRAS2011} and
references therein), offers a new opportunity to study Kerr spacetime by the
observation of various stars in inclined, highly elliptical orbits, and by the
analysis of their orbital dynamics
\cite{IorioPRD2011,SadeghianCQG2011,IorioMNRAS2011,WillAJL2008}. Relativistic
effects are difficult to discern since the orbital periods of the stars are in
the tens of years \cite{WillAJL2008}, and for orbits that come close to the
MBH, tidal disruption is a concern (\cite{WillAJL2008} and see Appendix B in
\cite{SadeghianCQG2011}); yet, observation has great potential to aid in the
study of Kerr spacetime. In the case of an EMRI, unfortunately, the part
played by a theoretician is a fiduciary one; thus, the introduction of tools
with which the evolution equations can be tested for consistency is most
beneficial: the abutment is one such tool, but it is not intended to replace
existing methods.

The concept of the abutment, a boundary that defines a set of near-polar
retrograde orbits, was developed and introduced by P. G. Komorowski in his
Doctoral thesis \cite{Komorowski:2011fk} and in a previous work
\cite{Komorowski:2010we} (we shall review the abutment in detail in section
\ref{sub:The-abutment.}); two uses of the abutment had emerged: first, it
suggested a means of testing the consistency of the evolution of the Carter
constant of circular orbits ($d{Q}/d{t}$) with respect to that of the latus
rectum ($d{\tilde{l}}/dt$); and second, it permitted a numerical analysis of
the rate of change of the orbital angle of inclination, $\iota$, with respect
to ${\tilde{l}}$ ($\left(  \partial{\iota}/\partial{\tilde{l}}\right)  _{\min
}$) for circular orbits constrained to evolve along the abutment. In this work
we shall extend these uses to orbits of non-zero eccentricity ($0\leq e\leq1$)
by testing the consistency of expressions for $d{Q}/d{t}$ with expressions for
$d{\tilde{l}}/dt$ and $d{e}/dt$, and we shall perform an analytical treatment
of $\iota$ and the list rate of the same. Further, a physically realistic
orbital evolution follows the abutment (${Q}_{X}$) in only one case, the
evolution of an orbit in a Schwarzschild black hole (SBH) system (${\tilde{S}%
}=0$). We shall now consider the general case of an evolving orbit that
intersects the abutment, ${Q}_{X}$, tangentially at a single point (contact of
the first order (see 99 in \cite{J.W.:1955fk})) as it follows a path defined
by ${Q}_{path}$. Further, by performing our analysis for elliptical orbits,
the abutment becomes a two dimensional surface that defines the maximum value
of ${Q}$ for given values of $e$ and latus rectum, ${\tilde{l}=l/M}$.
Therefore one must view the abutment as a set of contiguous points rather than
a path to be followed by an evolving orbit; and it is at these points that the
derivatives, $\partial{Q}_{X}/\partial{\tilde{l}}$ and $\partial{Q}%
_{X}/\partial{e}$, fix the corresponding slopes of ${Q}_{path}$. But as
reported in \cite{Komorowski:2010we}, the second-order effect\footnote{When we
refer to the second-order effect at the abutment, we refer to the second
derivative of $Q_{X}$ with respect to $\tilde{l}$, not to $\tilde{S}^{2}$.
\ See \cite{Komorowski:2010we} for background discussion.} must be included
when working with $\iota$ at the abutment.

In section \ref{sec:An-Analytical-Formula} we shall analytically derive the
formula for ${\iota}$ for elliptical orbits\ on the abutment, and thus confirm
the result for $\left(  \partial{\iota}/\partial{\tilde{l}}\right)  _{\min}$
\cite{Komorowski:2010we}, which was derived numerically for circular orbits.
In addition, we shall analytically derive $\partial{\iota}/\partial{e}$ for
elliptical orbits that evolve on the abutment. In section
\ref{sec:Correction-of-didl-for} we shall include the effect of the second
derivative of ${Q}_{path}$ (i.e. the second-order effect) by introducing
reductive ans\"{a}tze for circular and elliptical orbits, and thus create a more
physically realistic model for an evolving orbit at the abutment.

Because our abutment model is independent of any specific radiation
back-reaction model, we now have a laboratory that allows us to perform tests
of established listing formulae. In section
\ref{Treatment-of-Qdot-and-idot-at-the-abutment}, we shall demonstrate the
usefulness of the abutment in testing the consistency of $d{Q}/d{t}$ equations
with respect to $d\tilde{l}/dt$ and $de/dt$ evolution equations, and\ in
calculating $d\iota/dt$ for elliptical orbits of small eccentricity (i.e.
near-circular). In section \ref{Conclusions} we shall conclude our work and
recommend directions that warrant further study.

We define $\iota$ to be the maximum polar angle reached by the secondary
object in its orbit (see equation (42) in \cite{Komorowski:2010we}). This
definition differs from that used by others (Gair and Glampedakis
\cite{2006PhRvD..73f4037G} and Glampedakis, Hughes, and Kennefick
\cite{2002PhRvD..66f4005G}); but when performing our analysis to the leading
order in $\tilde{S}$, there is no significant difference.

\section{\label{sec:An-Analytical-Formula}An Analytical Formula for the Angle
of Inclination of an Elliptical Orbit on the Abutment}

\subsection{Introduction}

The listing of an inclined elliptical orbit of eccentricity ($e$) can be
described by $\partial{\iota}/\partial{\tilde{l}}$ and $\partial\iota/\partial
e$, where ${\iota}$ is the angle of inclination of the orbit and ${\tilde{l}}$
is its latus rectum normalised with respect to the mass (${M}$) of the Kerr
black hole (KBH). A set of essential analytical formulae for the orbital
constants of motion has been derived in \cite{Komorowski:2010we}: the Carter
constant at the abutment (${Q}_{X})$, the orbital energy (${\tilde{E}}$), and
the quantity, ${X}={\tilde{L}}_{z}-{\tilde{S}}{\tilde{E}}$, as well as an
analytical formula for ${\iota}$ in terms of these constants of motion.
Numerical analysis yielded an equation for $\left(  \partial{\iota}%
/\partial{\tilde{l}}\right)  _{\min}$ for circular orbits:%
\begin{align}
\left(  \frac{\partial{\iota}}{\partial{\tilde{l}}}\right)  _{min}  &
\cong-\left(  122.7{\tilde{S}}-36{\tilde{S}}^{3}\right)  {\tilde{l}}%
^{-11/2}-\left(  63/2{\tilde{S}}+35/4{\tilde{S}}^{3}\right)  {\tilde{l}%
}^{-9/2}\nonumber\\
&  -15/2{\tilde{S}}{\tilde{l}}^{-7/2}-9/2{\tilde{S}}{\tilde{l}}^{-5/2}.
\label{eq:didl_Laurent_III}%
\end{align}
To verify equation (\ref{eq:didl_Laurent_III}) analytically, we shall derive
the result to order 3 in ${\tilde{S}}$ (i.e. $O({\tilde{S}}^{3})$). Observe
that equation (\ref{eq:didl_Laurent_III}) is a series expansion in terms of
${\tilde{l}}^{-\frac{1}{2}}$. Further, the series coefficients are themselves
series expansions of odd powers of ${\tilde{S}}$. These are important
properties, which we shall confirm and investigate. Equation
(\ref{eq:didl_Laurent_III}) is not sufficient for understanding the effect of
radiation back-reaction on the listing of near-polar orbits; therefore, it is
necessary to develop an analytical formula for ${\iota}$ on the abutment so
that a more thorough treatment can be made. We shall review the analytical
formulae reported in \cite{Komorowski:2010we} for elliptical orbits, and
develop appropriate expansions of those formulae in terms of ${\tilde{S}}$.
The MacLaurin series expansions of the functions $1/(1+x)$, $\sqrt{1+x}$,
$\arccos(x)$, $\cos(x)$, and $\sin(x)$ are essential for this work.

\subsection{Review of Analytical Formulae}

\subsubsection{\label{sub:The-abutment.}The abutment, $Q_{X}$}

The analytical formula for ${X}_{\pm}^{2}$ (where ${X}={\tilde{L}}_{z}%
-{\tilde{S}}{\tilde{E}}$) for elliptical and inclined orbits about a KBH was
found to be \cite{Komorowski:2010we}:%
\begin{equation}
{X}_{\pm}^{2}={\frac{{Z}_{5}+{Z}_{6}{Q}\pm2{\tilde{S}}\,\sqrt{{Z_{7}}{Z_{8}%
}{Z_{9}}}}{\tilde{l}\left(  \tilde{l}\left(  3-\tilde{l}+{e}^{2}\right)
^{2}-4\,{\tilde{S}}^{2}\left(  1-{e}^{2}\right)  ^{2}\right)  }},
\label{eq:X2_III}%
\end{equation}
where
\begin{equation}
{Z}_{5}={\tilde{l}}^{3}\left\{  \left(  \tilde{l}+3\,{e}^{2}+1\right)
{\tilde{S}}^{2}-\tilde{l}\left(  3-\tilde{l}+{e}^{2}\right)  \right\}  ,
\label{Z5}%
\end{equation}%
\begin{equation}
{Z}_{6}=-2\,\left(  1-{e}^{2}\right)  ^{2}{\tilde{S}}^{4}+2\,\tilde{l}\left(
2\,{e}^{4}+\left(  2-\tilde{l}\right)  {e}^{2}+4-\tilde{l}\right)  {\tilde{S}%
}^{2}-{\tilde{l}}^{2}\left(  3-\tilde{l}+{e}^{2}\right)  ^{2}, \label{Z6}%
\end{equation}%
\begin{equation}
{Z_{7}}={\tilde{S}^{2}\left(  {1+{e}}\right)  ^{2}+\tilde{l}\left(  {\tilde
{l}-2(1+{e})}\right)  }, \label{Z7}%
\end{equation}%
\begin{equation}
{Z_{8}}={\tilde{S}^{2}\left(  {1-{e}}\right)  ^{2}+\tilde{l}\left(  {\tilde
{l}-2(1-{e})}\right)  }, \label{Z8}%
\end{equation}
and%
\begin{equation}
{Z_{9}}={\left(  {\tilde{l}^{5}+\tilde{S}^{2}Q^{2}\left(  {1-{e}^{2}}\right)
^{2}+Q\tilde{l}^{3}\left(  {3-\tilde{l}+{e}^{2}}\right)  }\right)  }.
\label{eq:Z9_III}%
\end{equation}
Intriguingly, the roots of ${Z_{7}=0}$ correspond to the coordinate
singularities associated with the event horizon of the KBH, multiplied by
$1+e$; for ${Z_{8}=0}$ the multiplier is $1-e$. The abutment, which lies
outside the event horizon, corresponds to a set of orbits for which ${Z}%
_{9}=0$ \cite{Komorowski:2010we}, i.e.%
\begin{equation}
{\tilde{l}}^{5}+{\tilde{S}}^{2}{Q}^{2}\left(  1-{e}^{2}\right)  ^{2}%
+{Q}{\tilde{l}}^{3}\left(  3-{\tilde{l}}+{e}^{2}\right)  =0.
\label{eq:LatusRectumGeneralAbutment_III}%
\end{equation}
The solution of equation (\ref{eq:LatusRectumGeneralAbutment_III}) is:%
\begin{equation}
{Q}_{X}=\frac{{\tilde{l}}^{2}}{2{\tilde{S}}^{2}{\left(  1-{e}^{2}\right)
}^{2}}\left(  \tilde{l}\left(  \tilde{l}-{e}^{2}-3\right)  \pm\sqrt{{\tilde
{l}}^{2}\left(  \tilde{l}-{e}^{2}-3\right)  ^{2}-4\,\tilde{l}\left(  1-{e}%
^{2}\right)  ^{2}{\tilde{S}}^{2}}\right)  . \label{eq:Q_Abutment_Analytical}%
\end{equation}
Where the minus solution must be taken since the plus solution has a
singularity at ${\tilde{S}=0}$ (unphysical for an SBH) and at $e=1$. Further,
the value of $Q$ of an evolving orbit cannot exceed $Q_{X}$; equation
(\ref{eq:X2_III}) would yield a complex result for ${X}_{\pm}^{2}$. Hence
$\tilde{L}_{z}$ and $\tilde{E}$ would possess unphysical values.

By performing an expansion in terms of ${\tilde{S}}^{2}$ one obtains:
\begin{equation}
Q_{X}\cong\frac{{\tilde{l}}^{2}}{\left(  {\tilde{l}}-{e}^{2}-3\right)  }%
+\frac{{\tilde{l}}\left(  1-{e}^{2}\right)  ^{2}{\tilde{S}}^{2}}{\left(
{\tilde{l}}-{e}^{2}-3\right)  ^{3}}+2\frac{\left(  1-{e}^{2}\right)
^{4}{\tilde{S}}^{4}}{\left(  {\tilde{l}}-{e}^{2}-3\right)  ^{5}}%
\ldots\label{eq:Q_Abutment_Intermediate_Form_2_III}%
\end{equation}
Therefore ${Q}_{X}=O\left(  {\tilde{S}}^{0}\right)  $ and the ${j}^{th}$ term
of ${Q}_{X}=O\left(  {\tilde{S}}^{2{j}}\right)  $. The expansion of $Q_{X}$ in
terms of $\tilde{l}$ can be derived from equation
(\ref{eq:Q_Abutment_Intermediate_Form_2_III}) once it has been determined to
which power of $\tilde{S}$ one wishes to work. This result, and its
derivatives with respect to $\tilde{l}$ and $e$, are presented in Appendix
\ref{Treatment-of-Qx-series} for use in our analysis in section
\ref{Treatment-of-Qdot-at-the-abutment}.

We return to equation (\ref{eq:X2_III}). The terms under the square root can
be excluded since ${Z}_{9}=0$. Substitution of ${Q}_{X}$ into the remaining
part of the equation yields:
\begin{align}
{X}_{\pm}^{2}  &  \cong{\tilde{S}}^{2}\biggl ({\frac{{\tilde{l}}\left(
{\tilde{l}}^{2}-4\,{\tilde{l}}-4\,{e}^{2}+4\right)  }{\left(  {\tilde{l}}%
-{e}^{2}-3\right)  ^{3}}}\label{eq:X2_S_Series_III}\\
&  +2\,{\frac{\left(  2-10\,{e}^{2}+{e}^{2}{\tilde{l}}-3\,{\tilde{l}}%
+{\tilde{l}}^{2}\right)  \left(  1-{e}^{2}\right)  ^{2}{\tilde{S}}^{2}%
}{\left(  {\tilde{l}}-{e}^{2}-3\right)  ^{5}}}\nonumber\\
&  +{\frac{\left(  6\,{\tilde{l}}^{2}+\left(  8\,{e}^{2}-16\right)  {\tilde
{l}}+9-74\,{e}^{2}+{e}^{4}\right)  \left(  1-{e}^{2}\right)  ^{4}{\tilde{S}%
}^{4}}{{\tilde{l}}\left(  {\tilde{l}}-{e}^{2}-3\right)  ^{7}}}%
\biggr ).\nonumber
\end{align}
From equation (\ref{eq:X2_S_Series_III}) one finds that ${X}_{\pm}%
^{2}=O\left(  {\tilde{S}}^{2},{\tilde{l}}^{0}\right)  $. Further analysis
yields the result:%
\begin{align}
{X}  &  =\pm{\tilde{S}}\sqrt{{\frac{\tilde{l}\left(  {\tilde{l}}^{2}%
-4\,\tilde{l}-4\,{e}^{2}+4\right)  }{\left(  \tilde{l}-{e}^{2}-3\right)  ^{3}%
}}}\label{eq:X_S_Series_III}\\
&  \times\mathopen\Bigg (1+{\frac{\left(  1-{e}^{2}\right)  ^{2}{\tilde{S}%
}^{2}}{\left(  {\tilde{l}}-{e}^{2}-3\right)  ^{2}\tilde{l}\left(  {\tilde{l}%
}^{2}-4\,\tilde{l}-4\,{e}^{2}+4\right)  }}{P}_{1}\nonumber\\
&  +\frac{1}{2}{\frac{\left(  1-{e}^{2}\right)  ^{4}{\tilde{S}}^{4}}{\left(
{\tilde{l}}-{e}^{2}-3\right)  ^{4}{\tilde{l}}^{2}\left(  {\tilde{l}}%
^{2}-4\,\tilde{l}-4\,{e}^{2}+4\right)  ^{2}}}{P}_{2}\Bigg)\nonumber
\end{align}
where%
\begin{align*}
{P}_{1}  &  ={\tilde{l}}^{2}-\left(  3-{e}^{2}\right)  \tilde{l}+2-10\,{e}%
^{2},\\
{P}_{2}  &  =5\,{\tilde{l}}^{4}-\left(  34-6\,{e}^{2}\right)  {\tilde{l}}%
^{3}+\left(  84-104\,{e}^{2}\right)  {\tilde{l}}^{2}\\
&  -\left(  88-328\,{e}^{2}+16\,{e}^{4}\right)  \tilde{l}+32-292\,{e}%
^{2}+200\,{e}^{4}-4\,{e}^{6}.
\end{align*}

\subsubsection{Orbital energy, ${\tilde{E}}$}

The formula for orbital energy, ${\tilde{E}}$, for inclined elliptical orbits
(see equation (44) in \cite{Komorowski:2010we}) is presented here in a form
that more clearly shows that ${\tilde{E}}=O\left(  {\tilde{S}}^{0}\right)  $:
\begin{equation}
\tilde{E}=\sqrt{1-{\left(  1-{e}^{2}\right)  }{\frac{{\tilde{l}}^{3}-Q\left(
\tilde{l}-{\tilde{S}}^{2}\right)  \left(  1-{e}^{2}\right)  -{\tilde{l}}%
{X}^{2}\left(  1-{e}^{2}\right)  }{{\tilde{l}}^{4}}}}; \label{eq:E4_III}%
\end{equation}
further, substitution of ${Q}_{X}\left(  {\tilde{S}}\right)  $ and ${X}_{\pm
}^{2}\left(  {\tilde{S}}\right)  $ into equation (\ref{eq:E4_III}) yields an
expression for ${\tilde{E}}$, which can be used directly in our analysis, or
in the following form:%
\begin{align}
\tilde{S}^{2}\left(  {1-\tilde{E}^{2}}\right)   &  =\frac{{\left(  {1-e^{2}%
}\right)  \left(  {{\tilde{l}}-4}\right)  {\tilde{S}}^{2}}}{{{\tilde{l}%
}\left(  {{\tilde{l}}-e^{2}-3}\right)  }}-2\,\frac{{\left(  {1-e^{2}}\right)
^{2}\left(  {e^{2}{\tilde{l}}+{\tilde{l}}-6\,e^{2}-2}\right)  {\tilde{S}}^{4}%
}}{{{\tilde{l}}^{2}\left(  {{\tilde{l}}-e^{2}-3}\right)  ^{3}}}\nonumber\\
&  -\frac{{\left(  {1-e^{2}}\right)  ^{4}\left(  {-3-30\,e^{2}+4\,e^{2}%
{\tilde{l}}+{\tilde{l}}^{2}+e^{4}}\right)  {\tilde{S}}^{6}}}{{{\tilde{l}}%
^{3}\left(  {{\tilde{l}}-e^{2}-3}\right)  ^{5}}}. \label{eq:S2E2_Series_III}%
\end{align}

\subsubsection{Orbital angle of inclination, ${\iota}$}

The exact formula for ${\iota}$ was derived in Boyer-Lindquist coordinates (BL
coordinates) and found to be:%
\begin{equation}
\sin^{2}\left(  {\iota}\right)  {\small =}\frac{{Q+\tilde{L}_{z}^{2}+\tilde
{S}^{2}\left(  {1-\tilde{E}^{2}}\right)  -\sqrt{\left(  {Q+\tilde{L}_{z}%
^{2}+\tilde{S}^{2}\left(  {1-\tilde{E}^{2}}\right)  }\right)  ^{2}%
-4\,Q\tilde{S}^{2}\left(  {1-\tilde{E}^{2}}\right)  }}}{2{\tilde{S}^{2}\left(
{1-\tilde{E}^{2}}\right)  }}, \label{eq:Theta-Roots-Equatorial_III}%
\end{equation}
which suggests an approximate expansion in the cases of small ${\tilde{S}}$ or
for ${\tilde{l}}\rightarrow\infty$ (for which ${\tilde{E}}\rightarrow1$)
\cite{Komorowski:2010we}. In particular, near-equatorial orbits can also be
approximated by such an expansion since ${Q}\gtrapprox0$. But we are studying
near-polar orbits, for which ${Q}>12$; so it is advantageous to exploit the
fact that ${\tilde{L}}_{z}\cong0$ and convert equation
(\ref{eq:Theta-Roots-Equatorial_III}) to an alternative form which can be
expanded as a series (not strictly in powers of ${\tilde{S}}^{2}$) to obtain:
\begin{align}
\cos^{2}\left(  {\iota}\right)   &  {\small \cong}\frac{{\tilde{L}}_{z}^{2}%
}{{Q+\tilde{L}_{z}^{2}-\tilde{S}^{2}\left(  {1-\tilde{E}^{2}}\right)  }%
}{\small -}\frac{{\tilde{L}}_{z}^{4}\left(  {1-\tilde{E}^{2}}\right)
\tilde{S}^{2}}{\left(  {Q+\tilde{L}_{z}^{2}-\tilde{S}^{2}\left(  {1-\tilde
{E}^{2}}\right)  }\right)  ^{3}}\nonumber\\
&  {\small +}\frac{2{\tilde{L}}_{z}^{6}\left(  {1-\tilde{E}^{2}}\right)
^{2}\tilde{S}^{4}}{\left(  {Q+\tilde{L}_{z}^{2}-\tilde{S}^{2}\left(
{1-\tilde{E}^{2}}\right)  }\right)  ^{5}} \label{eq:iota-series_III}%
\end{align}
It is essential to establish the lowest order of ${\tilde{S}}$ for each term
of equation (\ref{eq:iota-series_III}); the results in equations
(\ref{eq:Q_Abutment_Intermediate_Form_2_III}) and (\ref{eq:X2_S_Series_III}),
and equation (\ref{eq:S2E2_Series_III}) to $O\left(  \tilde{S}^{2}\right)
$\ will help.

It was found that ${X}_{\pm}=-\sqrt{{X}_{\pm}^{2}}$ in the vicinity of the
abutment (see section 3.5 in Komorowski et al. \cite{Komorowski:2010we});
therefore,
\begin{equation}
{\tilde{L}}_{z}=-\sqrt{{X}_{\pm}^{2}}+{\tilde{S}}{\tilde{E}}.
\label{eq:calc_Lz}%
\end{equation}
Each of the expressions in equations (\ref{eq:X_S_Series_III}) and
(\ref{eq:E4_III}), when expanded as a power series in $\tilde{l}^{-1}$, will
have a leading factor of $\tilde{S}$ and $unity$, respectively. In evaluating
equation (\ref{eq:calc_Lz}), the leading terms subtract out; therefore, we
find that ${\tilde{L}}_{z}^{2}=O\left(  {\tilde{S}}^{2},{\tilde{l}}%
^{-2}\right)  $. The inverse dependence of ${\tilde{L}}_{z}^{2}$\ on
$\tilde{l}$ is consistent with the physical meaning of ${\tilde{L}}_{z}$ for
orbits on the abutment. Further, equation
(\ref{eq:Q_Abutment_Intermediate_Form_2_III}) indicates that ${Q}_{X}=O\left(
{\tilde{S}}^{0},{\tilde{l}}\right)  $; therefore, the first term in equation
(\ref{eq:iota-series_III}) is $O\left(  {\tilde{S}}^{2}\right)  $, and the
second term, $O\left(  {\tilde{S}}^{6}\right)  $, with each term containing
higher order terms of $\tilde{S}$ in increments of $4$.

Taking the square root of both sides of equation (\ref{eq:iota-series_III})
yields,
\begin{equation}
\cos\left(  {\iota}\right)  ={\overset{\overset{1^{st}}{term}%
}{\overbrace{O\left(  {\tilde{S}}\right)  }}}+\overset{\overset{2^{nd}%
}{term}}{\overbrace{O\left(  \tilde{S}^{5}\right)  }}+\overset{\overset{3^{rd}%
}{term}}{\overbrace{O\left(  \tilde{S}^{9}\right)  }}
\label{eq:cos-iota-first-term}%
\end{equation}
with higher order terms of odd power of ${\tilde{S}}$. The second term in
equation (\ref{eq:iota-series_III}) will contribute to equation
(\ref{eq:cos-iota-first-term}) a factor $O\left(  {\tilde{S}}^{5}\right)  $;
therefore, to derive an analytical formula for $\left(  \partial{\iota
}/\partial{\tilde{l}}\right)  _{\min}$ valid to $O\left(  {\tilde{S}}%
^{3}\right)  $ (see equation (\ref{eq:didl_Laurent_III})) it is sufficient to
use the first term of equation (\ref{eq:iota-series_III}). If we choose to
work in stronger gravitational fields, for which terms of greater order in
${\tilde{S}}$ are required, then the second and possibly higher order terms in
equation (\ref{eq:iota-series_III}) would be used. But we wish to work with
terms that contain $\tilde{S}$ and $\tilde{S}^{3}$, to the exclusion of those
with $\tilde{S}^{5}$, so we shall restrict our analysis to the first term of
the series in equation (\ref{eq:iota-series_III}); after taking the square
root, it can be simplified to yield:
\begin{equation}
\cos\left(  {\iota}\right)  \cong\frac{{\tilde{L}}_{z}}{\sqrt{{Q}}}\left[
1-\frac{1}{2}\frac{\tilde{L}_{z}^{2}}{Q}+\frac{1}{2}\frac{\tilde{S}^{2}\left(
{1-\tilde{E}^{2}}\right)  }{Q}\right]  . \label{eq:cos-iota-third-order}%
\end{equation}
Given $x=\cos\left(  \iota\right)  $, one may calculate $\iota$ to $O\left(
\tilde{S}^{3}\right)  $ by using the MacLaurin series for $\arccos(x)$ to
$O\left(  {x}^{3}\right)  $.

\subsection{Analytical Formula for $\iota\left(  e,\tilde{l}\right)  $ on the
Abutment}

We shall now evaluate equation (\ref{eq:cos-iota-third-order}) analytically by
working with the constituent terms as series expansions in ${\tilde{S}}$, the
coefficients of which are expressed in terms of ${e}$ and ${\tilde{l}}$;\ the
result to third order in $\tilde{S}$ is our target. An apercu of the method by
which the expression in equation (\ref{eq:cos-iota-third-order}) is treated
appears in Appendix \ref{Apercu}.

\subsubsection{First-order in ${\tilde{S}}$}

To perform our calculation of $\iota$ to $O\left(  \tilde{S}\right)  $ (see
Appendix \ref{First_Order_Calculations}) it is sufficient to use:%
\begin{equation}
\overset{\left(  1\right)  }{\iota}=\frac{\pi}{2}-\overset{\left(  1\right)
}{\frac{{\tilde{L}}_{z}}{\sqrt{{Q}_{X}}}} \label{eq:iota_first_order_in_S}%
\end{equation}
where%
\begin{equation}
\overset{\left(  1\right)  }{\frac{{\tilde{L}}_{z}}{\sqrt{{Q}_{X}}}}%
=-{\tilde{S}}\left(  {e}^{2}+3\right)  \left(  {\frac{1}{{\tilde{l}}^{3/2}}%
}+{\frac{\left(  1+{e}^{2}\right)  }{{\tilde{l}}^{5/2}}}+{\frac{\left(
3+2\,{e}^{2}+{e}^{4}\right)  }{{\tilde{l}}^{7/2}}}+{\frac{\left(  9+5\,{e}%
^{2}+5\,{e}^{4}+{e}^{6}\right)  }{{\tilde{l}}^{9/2}}}\right)
\label{eq:LzQ_Series_First_Order_III}%
\end{equation}
and the number in parenthesis indicates the order in $\tilde{S}$ of the term
below it.

\subsubsection{Third-order in $\tilde{S}$}

Our third-order equations are more complicated. Consider the third-order
equation for $\iota$:%
\begin{equation}
\overset{\left(  3\right)  }{\iota}=\frac{\pi}{2}-x-\frac{1}{6}x^{3}
\label{eq:iota-third-order-in-S}%
\end{equation}
where
\begin{equation}
x=\overset{\left(  3\right)  }{\frac{{\tilde{L}}_{z}}{\sqrt{{Q}_{X}}}}\left(
1-\frac{1}{2}\left(  \overset{\left(  3\right)  }{\frac{{\tilde{L}}_{z}}%
{\sqrt{{Q}_{X}}}}\right)  ^{2}+\frac{1}{2}\overset{\left(  2\right)
}{\frac{\tilde{S}^{2}\left(  {1-\tilde{E}^{2}}\right)  }{{Q}_{X}}}\right)
\label{eq:x-cos-iota-second-order}%
\end{equation}
in which%

\begin{equation}
\overset{\left(  3\right)  }{\frac{{\tilde{L}}_{z}}{\sqrt{{Q}_{X}}}%
}=\overset{\left(  1\right)  }{\frac{{\tilde{L}}_{z}}{\sqrt{{Q}_{X}}}}%
-{\tilde{S}}^{3}\left(  1-{e}^{2}\right)  ^{2}\left(  \frac{1}{{\tilde{l}%
}^{7/2}}+\frac{1}{2}\,{\frac{11+5\,{e}^{2}}{{\tilde{l}}^{9/2}}}\right)
\label{eq:LzQ_Series_Third_Order_III}%
\end{equation}
and%

\begin{equation}
\overset{\left(  2\right)  }{\frac{\tilde{S}^{2}\left(  {1-\tilde{E}^{2}%
}\right)  }{{Q}_{X}}}=\left(  {1-{e}^{2}}\right)  \left(  {\frac{1}{{\tilde
{l}}^{2}}}-{\frac{4}{{\tilde{l}}^{3}}}\right)  {\tilde{S}}^{2}
\label{eq:S2E2Q_Series_III}%
\end{equation}
(see Appendix \ref{Third_Order_Calculations}). We evaluate equation
(\ref{eq:iota-third-order-in-S}) to obtain the final result, of $O\left(
\tilde{S}^{3}\right)  $:%

\begin{align}
\overset{\left(  3\right)  }{{\iota}}  &  =\left[  {\tilde{S}}^{3}\left(
-8-13\,{e}^{2}-2\,{e}^{4}+5/3\,{e}^{6}\right)  +\left(  {e}^{2}+3\right)
\left(  9+5\,{e}^{2}+5\,{e}^{4}+{e}^{6}\right)  \tilde{S}\right]  {\tilde{l}%
}^{-9/2}\nonumber\\
&  +\left[  1/2\,\left(  1-{e}^{2}\right)  \left(  5-{e}^{2}\right)
{\tilde{S}}^{3}+\left(  {e}^{2}+3\right)  \left(  3+2\,{e}^{2}+{e}^{4}\right)
\tilde{S}\right]  {\tilde{l}}^{-7/2}\nonumber\\
&  +\tilde{S}\left(  3+{e}^{2}\right)  \left(  1+{e}^{2}\right)  {\tilde{l}%
}^{-5/2}+\tilde{S}\left(  3+{e}^{2}\right)  {\tilde{l}}^{-3/2}{+\frac{\pi}%
{2}.} \label{eq:iota}%
\end{align}

\subsection{Derivatives of $\iota\left(  e,\tilde{l}\right)  $ on the
Abutment}

By taking the partial derivative of $\iota$ with respect to $\tilde{l}$ (using
equation (\ref{eq:iota}))\ one obtains:%
\begin{align}
\left(  \frac{\partial{\iota}}{\partial{\tilde{l}}}\right)  _{min}  &
=-\frac{3}{2}\left[  {3\,\left(  {e}^{2}+3\right)  \left(  9+5\,{e}^{2}%
+5\,{e}^{4}+{e}^{6}\right)  {\tilde{S}-}\left(  24+39\,{e}^{2}+6\,{e}%
^{4}-5\,{e}^{6}\right)  {\tilde{S}}^{3}}\right]  {\tilde{l}}^{-11/2}%
\nonumber\\
&  -\frac{7}{2}\left[  {\left(  {e}^{2}+3\right)  \left(  3+2\,{e}^{2}+{e}%
^{4}\right)  {\tilde{S}}+\frac{1}{2}\left(  1-{e}^{2}\right)  \left(
{5-e}^{2}\right)  {\tilde{S}}^{3}}\right]  {\tilde{l}}^{-9/2}\nonumber\\
&  -\frac{5}{2}\,{\left(  3+{e}^{2}\right)  \left(  1+{e}^{2}\right)  }%
{\tilde{S}}{\tilde{l}}^{-7/2}-\frac{3}{2}\,{\left(  3+{e}^{2}\right)  }%
{\tilde{S}}{\tilde{l}}^{-5/2}. \label{eq:didl_III}%
\end{align}
The partial derivative of $\iota$ with respect to $e$ can also be derived:%
\begin{align}
\left(  \frac{\partial{\iota}}{\partial{e}}\right)  _{min}  &  =2e\left(
4\left(  6+10\,{e}^{2}+6\,{e}^{4}+{e}^{6}\right)  \tilde{S}-\,\left(
13-5\,{e}^{4}+4\,{e}^{2}\right)  {\tilde{S}}^{3}\right)  {\tilde{l}}%
^{-9/2}\nonumber\\
&  +2e\left(  \left(  9+10\,{e}^{2}+3\,{e}^{4}\right)  \tilde{S}-\left(
3-{e}^{2}\right)  \tilde{S}^{3}\right)  {\tilde{l}}^{-7/2}\nonumber\\
&  +4e\left(  2+{e}^{2}\right)  \tilde{S}{\tilde{l}}^{-5/2}\,+2e\tilde
{S}{\tilde{l}}^{-3/2}\,{.} \label{eq:dide}%
\end{align}
The formula in equation (\ref{eq:didl_III}), when evaluated at $e=0$, matches
the numerical result in equation (\ref{eq:didl_Laurent_III}) for all of the
terms with the exception of $-122.7~\tilde{S}\tilde{l}^{-11/2}$, which differs
slightly from the analytical result of $-243/2~\tilde{S}\tilde{l}^{-11/2}$.

\subsection{Directional Derivatives in the $\tilde{l}-{e}$ Plane}

Consider the constant of motion, $Q$, and the corresponding quantity, $\iota$,
in the $\tilde{l}-e$ plane; by using the concept of the directional derivative
for two variables, one may represent $d{Q}/d{t}$ by the equation:%
\begin{equation}
\frac{{d{Q}}}{{d{t}}}=\frac{\partial{Q}}{\partial{\tilde{l}}}\frac{d{\tilde
{l}}}{dt}+\frac{\partial{Q}}{\partial{e}}\frac{d{e}}{dt}, \label{eq:Q_dot_III}%
\end{equation}
and in a similar manner we may define,
\begin{equation}
\frac{d{\iota}}{d{t}}=\frac{\partial{\iota}}{\partial{\tilde{l}}}%
\frac{d{\tilde{l}}}{dt}+\frac{\partial{\iota}}{\partial{e}}\frac{d{e}}{dt},
\label{eq:full_by_t}%
\end{equation}
where the terms $d{\tilde{l}/}dt$ and $d{e/}dt$ denote the evolution of
$\tilde{l}$ and $e$ to arbitrary order. We have the benefit of knowing the
analytical expressions $\partial Q/\partial\tilde{l}$ (see equation
(\ref{eq:dQxdl-appendix})) and $\partial Q/\partial e$ (see equation
(\ref{eq:dQxde-appendix})) at the abutment, which we can derive to the
required order.

A weak-field solution for $d\iota/dt$, in terms of $\tilde{l}$\ and $e$, has
been derived and reported in the literature (see equation (15a) in
\cite{Ryan:1996ly}):%
\begin{equation}
\frac{d\iota}{dt}=\frac{m\tilde{S}}{M^{2}}\tilde{l}^{-\frac{11}{2}}{\left(
1-{e}^{2}\right)  ^{\frac{3}{2}}}\sin\left(  \iota\right)  \left(  \frac
{244}{15}+\frac{252}{5}e^{2}+\frac{19}{2}e^{4}-\cos\left(  2\psi_{0}\right)
\left(  8e^{2}+\frac{26}{5}e^{4}\right)  \right)  , \label{eq:iota-dot-Ryan}%
\end{equation}
where the term $\cos\left(  2\psi_{0}\right)  $, in which $\psi_{0}$
represents the orientation of the elliptical orbit in the orbital plane,
typically averages to zero with the possible exception where the orbit has a
large value of $e<1$ \cite{Ryan:1996ly}. More recently, a solution for
$d\iota/dt$ to higher order in $\tilde{l}^{-1}$ (we present the weak-field
portion here) was derived by Flanagan and Hinderer \cite{Flanigan:2007kx}:%
\begin{equation}
\frac{d\iota}{dt}=\frac{m\tilde{S}}{M^{2}}\tilde{l}^{-\frac{11}{2}}{\left(
1-{e}^{2}\right)  }^{\frac{3}{2}}\sin\left(  \iota\right)  \left(  \frac
{266}{15}+\frac{184}{5}e^{2}+\frac{151}{20}e^{4}+\cos\left(  2\iota\right)
\left(  \frac{22}{15}-\frac{62}{5}e^{2}-\frac{39}{20}e^{4}\right)  \right)  ,
\label{eq:iota-dot-Flanagan-Hinderer}%
\end{equation}
in which they confirmed a weak-field correspondence to equation
(\ref{eq:iota-dot-Ryan}). In addition to the $\sin\left(  \iota\right)  $
contribution found in both equation (\ref{eq:iota-dot-Ryan}) and
(\ref{eq:iota-dot-Flanagan-Hinderer}), there is a $\cos\left(  2\iota\right)
$ term in the latter expression.

The trigonometric quantities, $\sin\left(  \iota\right)  $ and $\cos\left(
\iota\right)  $, do not occur in our expressions for $\iota$ and its
derivatives at the abutment. But such trigonometric terms are found, usually
in a product with $\tilde{S}$, in the general evolution equations (i.e.
$d\tilde{l}/dt$, $de/dt$, $dQ/dt$, and $d\iota/dt$) published in the
literature \cite{Barack:2004uq,2007PThPh.117.1041G,2007PhRvD..75b4005B}. One
may use equations (\ref{eq:iota_first_order_in_S}) and
(\ref{eq:LzQ_Series_First_Order_III}) to derive approximations of $\sin\left(
\iota\right)  $ and $\cos\left(  \iota\right)  $ suitable for working in the
leading order of $\tilde{S}$. Further, we may use the approximation of
$\cos\left(  2\iota\right)  $ to corroborate the conclusion that equation
(\ref{eq:iota-dot-Flanagan-Hinderer}) is the same as (\ref{eq:iota-dot-Ryan})
in the weak-field regime. These trigonometric approximations are only valid on
the abutment; thus, if it is necessary to perform a differentiation of a
trigonometric term (as in equation (\ref{eq:dYdi-appendix})), then the
differentiation must be performed before making the approximation. Such
limitations notwithstanding, the trigonometric approximations are of value to
us investigators since they afford us a systematic method for their treatment.

\section{\label{sec:Correction-of-didl-for}Correction of $\partial{\iota
}/\partial{\tilde{l}}$ and $\partial{\iota}/\partial{e}$ for Second-order
Effects}

\subsection{Introduction}

For circular orbits, Komorowski et al. \cite{Komorowski:2010we} found that the
numerical estimate of $\left(  \partial{\iota}/\partial{\tilde{l}}\right)
_{\min}$ in the weak-field regime deviates from the $\partial{\iota}%
/\partial{\tilde{l}}$ results reported in the literature (see Flanagan and
Hinderer \cite{Flanigan:2007kx}, and Hughes \cite{PhysRevD.61.084004}).
Consider the quotient of the formulae presented in equation (3.9) of Hughes
\cite{PhysRevD.61.084004} where ${\iota}\cong\pi/2$:
\begin{equation}
\frac{{\dot{\iota}}_{weak}}{{\dot{R}}_{weak}}=\frac{\partial{\iota}}%
{\partial{\tilde{l}}}=-{\frac{61}{48}}{\tilde{S}}{\tilde{l}}^{-\frac{5}{2}}.
\label{eq:Weak-RR_III}%
\end{equation}
Because $-61/48>-4.5$ in the weak-field regime, ${X}_{+}^{2}\Rightarrow{X}%
_{-}^{2}$ is the pertinent mode; and the best information one can obtain from
$\left(  \partial{\iota}/\partial{\tilde{l}}\right)  _{min}$ is the
specification of the lower limit of $\partial{\iota}/\partial{\tilde{l}}$ for
all ${\tilde{l}}>{\tilde{l}}_{LSO,\,abutment}$ (see \cite{Komorowski:2010we}
for more details about the abutment and its relationship to the last stable
orbit (LSO)). Therefore the second-order (i.e. $\partial^{2}{Q}_{path}%
/\partial{\tilde{l}}^{2}$) behaviour at the point of tangential intersection
of ${Q}_{X}$ and ${Q}_{path}$ must be considered. In section
\ref{sec:An-Analytical-Formula} the numerical results have been verified by
analytical derivation of the formula for $\left(  \partial{\iota}%
/\partial{\tilde{l}}\right)  _{\min}$ to $O\left(  \tilde{S}^{3}\right)  $. It
remains for us to extend this analysis to include second-order effects on
elliptical orbits;\ to this end, we shall discuss how to incorporate
second-order effects into $Q_{X}$, and the resultant change to the formula for
$X_{\pm}^{2}$ (see equation (\ref{eq:X2_III})). Equation
(\ref{eq:iota_first_order_in_S}) is sufficient in treating ${X}_{\pm}^{2}$,
and then ultimately $\iota$, to the leading order in $\tilde{S}$.

\subsection{Second-order Effects in ${Q}_{path}$}

\subsubsection{Circular orbits}

Let us begin our treatment in the ${Q}-{\tilde{l}}$ plane with the value of
$e$ held constant at zero. The form of ${Q}_{X}$ is represented by the series
in equation (\ref{eq:Q_Abutment_Intermediate_Form_2_III}); and because
${Q}_{path}$ intersects ${Q}_{X}$ tangentially at a single point (${\tilde{l}%
}_{o}$) (contact of the first order), we surmise:
\begin{equation}
{Q}_{path}\mathopen\Bigg |_{{\tilde{l}}={\tilde{l}}_{o}}={Q}_{X}%
\mathopen\Bigg |_{{\tilde{l}}={\tilde{l}}_{o}} \label{eq:Qpath=QX}%
\end{equation}
and
\begin{equation}
\frac{\partial{Q}_{path}}{\partial{\tilde{l}}}\mathopen\Bigg |_{{\tilde{l}%
}={\tilde{l}}_{o}}=\frac{\partial{Q}_{X}}{\partial{\tilde{l}}}%
\mathopen\Bigg |_{{\tilde{l}}={\tilde{l}}_{o}}. \label{eq:dQpath=dQX}%
\end{equation}
But the abutment can only offer an upper bound on the second derivative of
${Q}_{path}$, i.e.
\begin{equation}
\frac{\partial^{2}{Q}_{path}}{\partial{\tilde{l}}^{2}}%
\mathopen\Bigg |_{{\tilde{l}}={\tilde{l}}_{o}}\leqq\frac{\partial^{2}{Q}_{X}%
}{\partial{\tilde{l}}^{2}}\mathopen\Bigg |_{{\tilde{l}}={\tilde{l}}_{o}}.
\label{eq:Upper-Bound-Second-Derivative}%
\end{equation}
To perform an analytical treatment of the second derivative of ${Q}_{path}$,
we define an ansatz:%
\begin{equation}
{Q}_{path}={Q}_{X}-\frac{{\lambda}^{2}}{2}f\left(  {\tilde{l}}_{o}\right)
\label{eq:The_Reductive_Ansats}%
\end{equation}
where
\begin{equation}
f\left(  {\tilde{l}}_{o}\right)  =\left(  {\tilde{l}}_{o}\right)  ^{p}\left(
{\tilde{S}}\right)  ^{q}\left(  \sum_{k=0}^{n}{a}_{k}\left(  {\tilde{l}}%
_{o}\right)  ^{-k}\right)  , \label{eq:Reductive_Ansats_f}%
\end{equation}%
\[
{a}_{0}>0,
\]
and
\begin{equation}
{\lambda}={\tilde{l}}-{\tilde{l}}_{o}, \label{eq:Reductive_Ansats_lambda}%
\end{equation}
where ${p}$ and ${q}$ shall be determined by requiring that the weak-field
solution be of the form, ${\tilde{S}}{\tilde{l}}^{-\frac{5}{2}}$ (see
equations (\ref{eq:didl_III}) and (\ref{eq:Weak-RR_III})). The adjustment
represented by equation (\ref{eq:The_Reductive_Ansats}) is based on the Taylor
expansion of a function; the function ${f\left(  {\tilde{l}}_{o}\right)  }$
represents a second derivative of a primitive, $\wp(e,\tilde{l})$, with
respect to $\tilde{l}$, which is evaluated at ${\tilde{l}}_{o}$. One must not
confuse the concepts of the abutment and Taylor series; equation
(\ref{eq:The_Reductive_Ansats}) is not intended to be a Taylor series
representation of $Q_{path}$. We have taken the analytical formula for $Q_{X}$
and incorporated a term, which is designed to adjust the second derivative of
$Q_{path}$ so that it makes contact with $Q_{X}$ tangentially at a prescribed
point, $\tilde{l}_{o}$. If $\tilde{l}=\tilde{l}_{o}$, then the adjustment to
$Q_{path}$ and $\partial Q_{path}/\partial\tilde{l}$ is zero; and the value of
$\partial^{2}Q_{path}/\partial\tilde{l}^{2}$\ is reduced by $f\left(
\tilde{l}_{o}\right)  $. Equation (\ref{eq:The_Reductive_Ansats}) can be
applied to the analytical development of $\partial{\iota}/\partial{\tilde{l}}%
$. We shall call this reduction of the second derivative the reductive ansatz circular.

Let us consider how the reductive ansatz circular affects equation
(\ref{eq:X2_III}), with attention given to equations (\ref{eq:Z9_III}) and
(\ref{eq:LatusRectumGeneralAbutment_III}); although we begin with an analysis
of circular orbits, $e$ has been retained in these equations for later use.
Evaluate
\begin{align}
Z_{9}\left(  Q_{path}\right)   &  =Z_{9}(Q_{X}-\Phi)\nonumber\\
&  =\left[  {\tilde{l}}^{5}+{\tilde{S}}^{2}\left(  1-e^{2}\right)  ^{2}%
Q_{X}^{2}-\left(  {\tilde{l}}-3-e^{2}\right)  {\tilde{l}}^{3}Q_{X}\right]
\nonumber\\
&  +\Phi\left(  2{\tilde{S}}^{2}\left(  1-e^{2}\right)  ^{2}Q_{X}+\Phi
{\tilde{S}}^{2}\left(  1-e^{2}\right)  ^{2}+\left(  {\tilde{l}}-3-e^{2}%
\right)  {\tilde{l}}^{3}\right)  \label{eq:Z9_Qpath_square_bracket_zero}%
\end{align}
where
\[
\Phi=\frac{{\lambda^{2}}}{2}{f\left(  {\tilde{l}}_{o}\right)  ,}%
\]
for which the quantity in square brackets in equation
(\ref{eq:Z9_Qpath_square_bracket_zero}) is equal to zero (\textit{viz.}
equation (\ref{eq:LatusRectumGeneralAbutment_III})) for all values of
${\tilde{l}}>{\tilde{l}}_{LSO,\,abutment}$; therefore, the use of this
reductive ansatz has assured us of an effective means to simplify the
expressions. The terms that remain share a common factor, $\lambda^{2}$, which
will appear as $\pm\lambda$ when taken outside of the square root in equation
(\ref{eq:X2_III}). We shall limit our analysis to $O\left(  \tilde{S}\right)
$ (the $\tilde{S}^{2}$ terms will affect terms of higher order in $\tilde
{l}^{-1}$ in the series in equations (\ref{eq:didl_III}) and (\ref{eq:dide}));
therefore, the product of $Z_{7}$ (equation (\ref{Z7})), $Z_{8}$ (equation
(\ref{Z8})), and $Z_{9}$ (equation(\ref{eq:Z9_Qpath_square_bracket_zero}))
simplifies to:
\begin{align}
Z_{\sqrt{\bullet}}  &  =Z_{7}Z_{8}Z_{9}\nonumber\\
&  =\Phi\tilde{l}^{5}\left(  \tilde{l}-2\left(  1+e\right)  \right)  \left(
\tilde{l}-2\left(  1-e\right)  \right)  \left(  \tilde{l}-3-e^{2}\right)  ,
\label{eq:Zsquareroot}%
\end{align}
where we evaluate $Z_{\sqrt{\bullet}}$ at the point of intersection on the
abutment by setting ${\tilde{l}}_{o}=\tilde{l}$ (i.e. $\lambda=0$). We take
the square root of $Z_{\sqrt{\bullet}}$, and a term, $\pm\lambda\sqrt{2}/2$,
emerges. The choice of sign is determined by the mode at the abutment.

\begin{figure}[ptb]%
\centering
\includegraphics[scale=0.2]{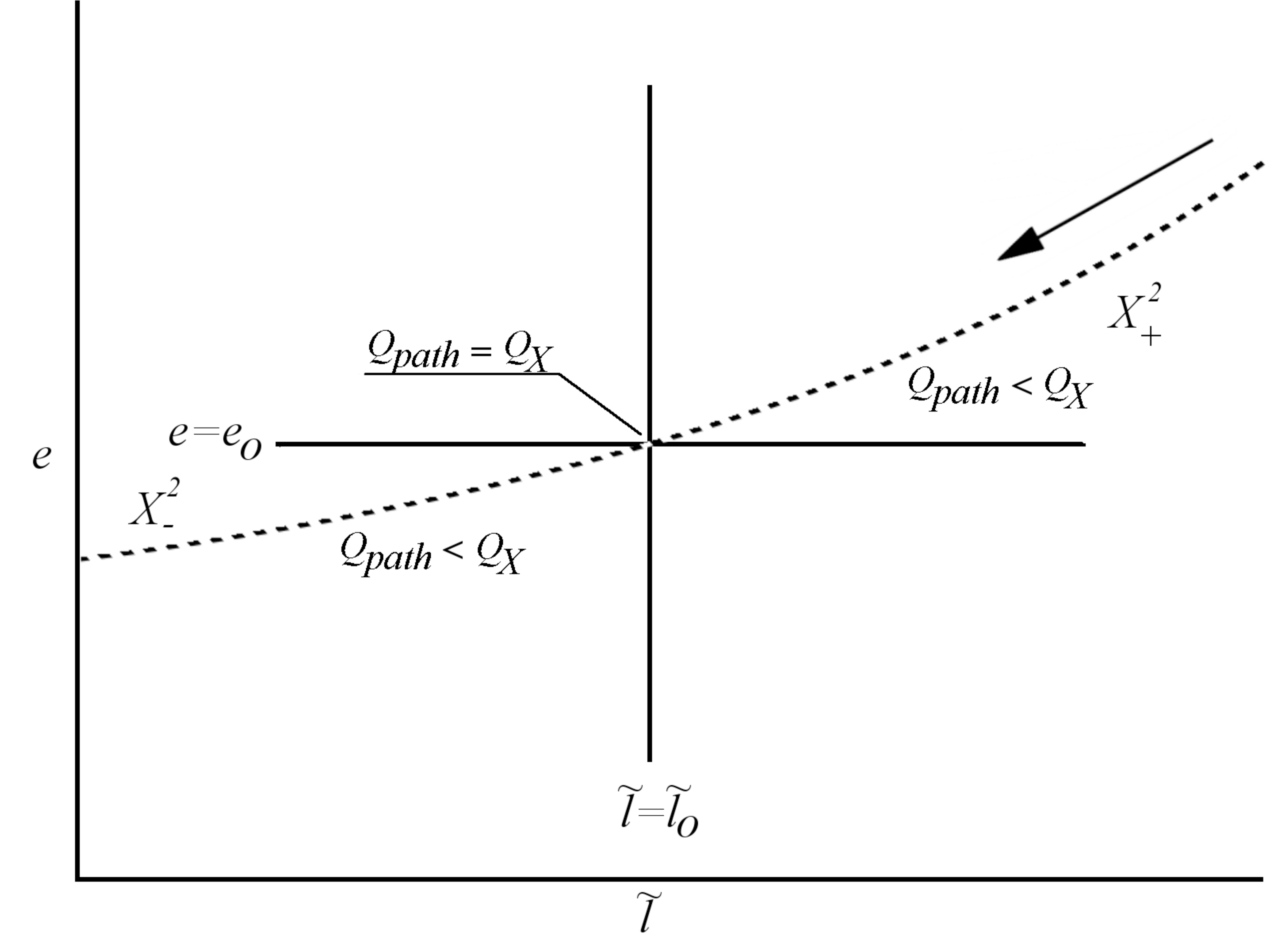}
\caption{A generic representation of $Q_{path}$ (short-dashed line) is
depicted in the $\tilde{l}-e$ plane as it makes contact of the first order at
a point $(\tilde{l}_{o},e_{o})$ on the $Q_{X}$ surface. \ The direction in
which the orbit evolves is shown by the arrow. \ We offer a generic
representation to emphasise that any $Q_{path}$, which is predicted by a
radiation back-reaction model, may be tested in this manner.}%
\label{fig:mode-map}%
\end{figure}

Mathematically speaking there are two modes at the abutment: the fast mode%
\begin{equation}
{X}_{-}^{2}\Rightarrow{X}_{+}^{2}, \label{eq:Mode_Fast_III}%
\end{equation}
and the slow mode
\begin{equation}
{X}_{+}^{2}\Rightarrow{X}_{-}^{2}. \label{eq:Mode_Slow_III}%
\end{equation}
In section 5 of \cite{Komorowski:2010we} it was established that orbits that
evolve on a path towards the abutment (during which $\tilde{l}>\tilde{l}_{o}$
and $\lambda>0$) are governed by ${X}_{+}^{2}$ (see equation (\ref{eq:X2_III}%
)) and after making contact with the abutment at $\tilde{l}=\tilde{l}_{o}$ the
orbits are then governed by ${X}_{-}^{2}$ (for which $\tilde{l}<\tilde{l}_{o}$
and $\lambda<0$) (see figure \ref{fig:mode-map}). Thus by choosing the
positive sign for $\pm\lambda$ the equation remains consistent with the
dominance of the slow mode. If one were to perform an analysis for the fast
mode then $-\lambda$ would be used instead.

An examination of equations (\ref{Z5}), (\ref{Z6}), and
(\ref{eq:Q_Abutment_Intermediate_Form_2_III}) reveals that
\begin{equation}
Z_{5}+Z_{6}Q_{X}=O\left(  \tilde{S}^{2},{\tilde{l}}^{4}\right)  ,
\label{eq:Order_of_Z5Z6Q}%
\end{equation}
\ from which one may infer%
\begin{equation}
2\tilde{S}\sqrt{Z_{\sqrt{\bullet}}}=O\left(  \tilde{S}^{2}\right)
\end{equation}
$\Rightarrow$%
\begin{equation}
\sqrt{2}\lambda\tilde{S}\times\tilde{S}^{q/2}=O\left(  \tilde{S}^{2}\right)  ;
\end{equation}
therefore, $q=2$ in the reductive ansatz (see equation
(\ref{eq:Reductive_Ansats_f})). The value of $p$ can be derived by considering
the order of $\tilde{L}_{z}$ in $\tilde{l}$. We find (\textit{viz.} equation
(\ref{eq:calc_Lz})) that $\tilde{L}_{z}=O\left(  \tilde{l}^{-1}\right)  $,
which must not be changed by the reductive ans\"{a}tze. And the leading term,
$\tilde{S}$, in the expression for $X$ (see equation (\ref{eq:X_S_Series_III}%
)) must remain. Given the order of $\tilde{l}$ in equation
(\ref{eq:Order_of_Z5Z6Q}), one must work with the next lower order, i.e.,
\begin{equation}
2\tilde{S}\sqrt{Z_{\sqrt{\bullet}}}=O\left(  {\tilde{l}}^{3}\right)
\end{equation}
$\Rightarrow$%
\begin{equation}
\pm\frac{\sqrt{2}}{2}\lambda\tilde{l}^{p/2}\tilde{l}^{4}=O\left(  {\tilde{l}%
}^{3}\right)  .
\end{equation}
Given $\lambda=O\left(  {\tilde{l}}\right)  $, we conclude that $p=-4$. In our
reductive ans\"{a}tze, we have found the values of $p$ and $q$ that ensure the
second-order effect does not change the form of $\partial\iota/\partial
\tilde{l}$ in the weak-field regime.

\subsubsection{\label{Common-Primative}Elliptical orbits}

The general formulation of the reductive ansatz elliptical can be derived by
starting with a Taylor series for two variables (see Appendix
\ref{Taylors_Series_for_two_Variables}). Because we concern ourselves with
second-order effects, we shall use the following operator:%
\begin{equation}
\frac{1}{2!}\left(  \lambda\frac{\partial}{\partial{\tilde{l}}}+\epsilon
\frac{\partial}{\partial{e}}\right)  ^{2}
\label{eq:Taylor_series_second_derivative_operator}%
\end{equation}
where $\lambda=(\tilde{l}-{\tilde{l}}_{o})$ and $\epsilon=\left(
e-e_{o}\right)  $, and where the ordered pair $\left(  e_{o},{\tilde{l}}%
_{o}\right)  $ specifies the location of the contact of first order between
$Q_{path}$ and $Q_{X}$ (see figure \ref{fig:mode-map}).

One may define the reductive ansatz elliptical, i.e.%
\begin{align}
Q_{path}  &  =Q_{X}-\frac{1}{2}\left[  \left(  \lambda\frac{\partial}%
{\partial\tilde{l}}+\epsilon\frac{\partial}{\partial e}\right)  ^{2}%
\wp(e,\tilde{l})\right]  _{{e}=e_{o},{\tilde{l}=\tilde{l}_{o}}}\nonumber\\
&  =Q_{X}-\frac{\lambda^{2}}{2}\left[  \frac{\partial^{2}}{\partial\tilde
{l}^{2}}\wp(e,\tilde{l})\right]  _{{e}=e_{o},{\tilde{l}=\tilde{l}_{o}}%
}-\lambda\epsilon\left[  \frac{\partial^{2}}{\partial\tilde{l}\partial e}%
\wp(e,\tilde{l})\right]  _{{e}=e_{o},{\tilde{l}=\tilde{l}_{o}}}-\frac
{\epsilon^{2}}{2}\left[  \frac{\partial^{2}}{\partial e^{2}}\wp(e,\tilde
{l})\right]  _{{e}=e_{o},{\tilde{l}=\tilde{l}_{o}}}
\label{eq:Two_Variable_Reductive_Ansatz}%
\end{align}
where we conjecture the existence of a primitive function, $\wp(e,\tilde{l})$.
Equation (\ref{eq:Upper-Bound-Second-Derivative}) will also be applied to the
case of elliptical orbits where $e$ is close to zero.

The expression for $Q_{path}$ is best regarded as a parameterized curve, and
to make such a treatment in equation (\ref{eq:Two_Variable_Reductive_Ansatz}),
one may factor out the $\lambda$, to obtain%
\begin{align}
Q_{path}  &  =Q_{X}-\frac{\lambda^{2}}{2}\left\{  \left[  \frac{\partial^{2}%
}{\partial\tilde{l}^{2}}\wp(e,\tilde{l})\right]  _{{e}=e_{o},{\tilde{l}%
=\tilde{l}_{o}}}+2\left(  \frac{\epsilon}{\lambda}\right)  \left[
\frac{\partial^{2}}{\partial\tilde{l}\partial e}\wp(e,\tilde{l})\right]
_{{e}=e_{o},{\tilde{l}=\tilde{l}_{o}}}+\left(  \frac{\epsilon}{\lambda
}\right)  ^{2}\left[  \frac{\partial^{2}}{\partial e^{2}}\wp(e,\tilde
{l})\right]  _{{e}=e_{o},{\tilde{l}=\tilde{l}_{o}}}\right\} \nonumber\\
&  =Q_{X}-\frac{\lambda^{2}}{2}g\left(  e_{o},\tilde{l}_{o}\right)  ,
\label{eq:Reductive-Ansatz-Elliptical}%
\end{align}
for which we have the benefit of knowing the limiting form of $\epsilon
/\lambda$ ($=de/d\tilde{l}$) to arbitrary order in $\tilde{l}^{-1}$. Thus it
is possible to parameterize $Q_{path}$\ in terms of $\lambda$. We can use the
expression,%
\begin{equation}
g\left(  e_{o},\tilde{l}_{o}\right)  ={\tilde{S}}^{2}\sum_{i=0}^{n}\frac
{{a}_{i}(e_{o})}{{\tilde{l}}_{o}^{i+4}};
\end{equation}
and this will form the basis of the reductive ansatz elliptical.

\subsection{Application of the Reductive Ans\"atze to the Analytical
Derivation of $\partial{\iota}/\partial{\tilde{l}}$ and $\partial{\iota
}/\partial{e}$}

The reductive ans\"atze (equations (\ref{eq:The_Reductive_Ansats}%
-\ref{eq:Reductive_Ansats_lambda}) and (equation
(\ref{eq:Reductive-Ansatz-Elliptical})) constitute a reduction of the second
derivative of ${Q}_{X}$ to more realistically model the behaviour of
${Q}_{path}$ at the abutment and perform a methodical treatment of this
reduction in the analytical calculation of $\partial{\iota}/\partial{\tilde
{l}}$ and $\partial{\iota}/\partial{e}$.

The procedure outlined in Appendix
\ref{Generalised_Treatment_of_the_Taylors_Series} yields the following formula
for $\partial{\iota}/\partial{\tilde{l}}$:%
\begin{equation}
\frac{\partial{\iota}}{\partial{\tilde{l}}}=\frac{\partial\left[
{\iota\left(  e,{\tilde{l}},{\lambda},{\tilde{S}}\right)  }\right]
_{{\lambda}=0\,}}{\partial{\tilde{l}}}+\left[  \frac{\partial\left[
{\iota\left(  e,{\tilde{l}},{\lambda},{\tilde{S}}\right)  }\right]  }%
{\partial{\lambda}}\frac{\partial{\lambda}}{\partial{\tilde{l}}}\right]
_{{\lambda}=0}, \label{eq:Total_derivative_wrt_l_chain}%
\end{equation}
where
\[
\frac{\partial{\lambda}}{\partial{\tilde{l}}}=1\text{ and }{\lambda
}\mathopen\Bigg |_{{\tilde{l}}={\tilde{l}}_{o}}=0;
\]
but the result for $\partial\iota/\partial e$ is simpler,%
\begin{align}
\frac{\partial{\iota}}{\partial{e}}  &  =\frac{\partial\left[  {\iota\left(
e,{\tilde{l}},{\lambda},{\tilde{S}}\right)  }\right]  _{{\lambda}=0}}%
{\partial{e}}\nonumber\\
&  =\left(  \frac{\partial\iota}{\partial e}\right)  _{\min}.
\label{eq:Total_derivative_wrt_e_chain}%
\end{align}

Equation (\ref{eq:Total_derivative_wrt_e_chain}) and the first term in
equation (\ref{eq:Total_derivative_wrt_l_chain}) yield the formulae that
describe the evolution of $\iota$ for a $Q_{path}$ along the abutment (i.e.
$\left(  \partial\iota/\partial e\right)  _{\min}$ and $(\partial{\iota
}/\partial{\tilde{l})}_{\min}$). The second term of equation
(\ref{eq:Total_derivative_wrt_l_chain}) incorporates second-order effects, and
thus describes the physically more realistic situation in which $Q_{path}$
intersects the abutment tangentially at a single point. Because one takes the
first derivative with respect to $\lambda$, the second and higher powers of
$\lambda$ will vanish when setting $\lambda=0$. But as we shall presently see,
the second-order effects of the reductive ans\"{a}tze remain.

\ We choose to work with the symbols $e$ and ${\tilde{l}}$ rather than $e_{o}$
and ${\tilde{l}}_{o}$, given that $e$ and ${\tilde{l}}$ can be used to
represent an arbitrary point on the abutment (see Appendix
\ref{Generalised_Treatment_of_the_Taylors_Series}). The reductive ansatz
circular (see equations (\ref{eq:The_Reductive_Ansats}%
-\ref{eq:Reductive_Ansats_lambda})) is applied at the abutment with $p=-4$,
$q=2$, and $n=2$ (while retaining the two terms of leading order in $\tilde
{S}$\ at the conclusion of the calculation) with%
\begin{equation}
\frac{\partial\iota}{\partial\tilde{l}}=-\tilde{S}\left\{  \frac{15}{2}%
+\frac{3}{2}A_{0}(0)-\frac{1}{4}A_{1}(0)\right\}  \tilde{l}^{-7/2}-\tilde
{S}\left\{  \frac{9}{2}-A_{0}(0)\right\}  \tilde{l}^{-5/2}
\label{eq:didl_Reduced}%
\end{equation}
where%
\[
A_{0}(e)=+\frac{\sqrt{2}}{2}\sqrt{a_{0}(e)},\text{\ and\ }A_{1}\left(
e\right)  =\frac{a_{1}(e)}{A_{0}(e)}.
\]

To apply this method to elliptical orbits, we will be required to use
$g\left(  e,\tilde{l}\right)  $. To calculate that function, the common
primitive $\wp(e,\tilde{l})$ is needed.

\subsection{Analytical Derivation of the Common Primitive, $\wp(e,\tilde{l})
$}

Now that the values of the parameters, $p=-4$ and $q=2$, have been found, it
is possible to derive the formula for $\wp(e,\tilde{l})$. Consider the
reductive ansatz circular:
\begin{equation}
f\left(  {\tilde{l}}\right)  ={\tilde{S}}^{2}\sum_{i=0}^{n}\frac{{a}_{i}%
}{{\tilde{l}}^{i+4}}.
\end{equation}
We conjecture a more general form of $f\left(  {\tilde{l}}\right)  $ that
includes $e$:%
\begin{equation}
f\left(  e,{\tilde{l}}\right)  ={\tilde{S}}^{2}\sum_{i=0}^{n}\frac{{a}%
_{i}\left(  1+b_{i}e^{2}\right)  }{{\tilde{l}}^{i+4}}.
\end{equation}
Performing the first integration over $\tilde{l}$ yields:%
\begin{align}
F\left(  e,\tilde{l}\right)   &  =\int f\left(  e,\tilde{l}\right)  d\tilde
{l}\nonumber\\
&  =-\tilde{S}^{2}\left[  \sum_{i=0}^{n}\left(  \frac{1}{i+3}\frac{{a}%
_{i}\left(  1+b_{i}e^{2}\right)  }{{\tilde{l}}^{i+3}}\right)  -\varkappa
\left(  e\right)  \right]  .
\end{align}
The second integration over $\tilde{l}$ yields an expression for the common
primitive:%
\begin{align}
\wp(e,\tilde{l})  &  =\int F\left(  e,\tilde{l}\right)  d\tilde{l}\nonumber\\
&  =\tilde{S}^{2}\left[  \sum_{i=0}^{n}\left(  \frac{1}{\left(  i+2\right)
\left(  i+3\right)  }\frac{{a}_{i}\left(  1+b_{i}e^{2}\right)  }{\tilde
{l}^{i+2}}\right)  +\varkappa\left(  e\right)  \tilde{l}+\zeta\left(
e\right)  \right]  .
\end{align}

The constants of integration, $\varkappa\left(  e\right)  \tilde{l}$ and
$\zeta\left(  e\right)  $, can each be set to zero since we require
$\lim\limits_{\tilde{l}\rightarrow\infty}\wp(e,\tilde{l})=0$. Now that the
formula for $\wp(e,\tilde{l})$ is known, it is possible to obtain
$g(e,\tilde{l})$, which is required by the reductive ansatz elliptical.

\section{\label{Treatment-of-Qdot-and-idot-at-the-abutment}The Treatment of
$d{Q}/d{t}$ and $d{\iota}/d{t}$ on the Abutment}

\subsection{\label{Treatment-of-Qdot-at-the-abutment}The $d{Q}/d{t}$ Evolution
Equations}

Komorowski et al. \cite{Komorowski:2010we} investigated the consistency of
$d{Q}/d{t}$ with the evolution equation $d\tilde{l}/dt$, for circular orbits
at the abutment ($\iota\gtrapprox\pi/2)$ by performing a preliminary numerical
analysis for values of $\tilde{l}=\left\{  7.0,100.0\right\}  $ and KBH spin
$\tilde{S}=\left\{  0.05,0.95\right\}  $ (see section 5.2.1 of
\cite{Komorowski:2010we}). The published values of $d\tilde{l}/dt$
\cite{PhysRevD.61.084004}, which we used in our investigation, were calculated
for $\iota\simeq\pi/3$, and the difference of this value of $\iota$ from that
at the abutment contributed to some inaccuracy in the analysis
\cite{Komorowski:2010we}. In this work, the derivation of analytical formulae
for $\iota$ and its derivatives, as well as the use of the directional
derivative to determine $d{Q}/d{t}$, now allow one to perform a more complete
treatment for elliptical orbits.

Let us consider the directional derivative in equation (\ref{eq:Q_dot_III}) as
a means of deriving $d{Q}/d{t}$ at the abutment. We have demonstrated that the
second-order effects are not seen when calculating the first derivatives of
$Q_{path}$ (i.e. $\partial Q_{path}/\partial\tilde{l}$ and $\partial
Q_{path}/\partial e$, see equation (\ref{eq:dQpath=dQX}) or
(\ref{eq:Reductive-Ansatz-Elliptical})); therefore, we may use $\partial
Q_{X}/\partial\tilde{l}$ (equation (\ref{eq:dQxdl-appendix})) and $\partial
Q_{X}/\partial e$ (equation\ (\ref{eq:dQxde-appendix})) when working with
equation (\ref{eq:Q_dot_III}).

The form of $dQ/dt$ (equation (A.3) in \cite{2007PhRvD..76d4007B} (after
equation (56) in \cite{2006PhRvD..73f4037G})), which was used in
\cite{Komorowski:2010we} to test $dQ/dt$ will be revisited in this work: \
\begin{align}
\left(  \frac{d{Q}}{d{t}}\right)  _{2PN}  &  =-\left(  1-\frac{1}{2}%
\frac{{\tilde{S}}^{2}\left(  3+e^{2}\right)  ^{2}}{\tilde{l}^{3}}\right)
\frac{64}{5}\frac{{m}^{2}}{M}{\left(  1-{e}^{2}\right)  ^{3/2}}\frac{\sqrt
{{Q}}}{{\tilde{l}}^{7/2}}\nonumber\\
&  \times\Bigl [{g}_{9}-\frac{{g}_{11}}{{\tilde{l}}^{1}}+{\pi}\frac{{g}_{12}%
}{{\tilde{l}}^{3/2}}-\frac{\left(  {g}_{13}-{\tilde{S}}^{2}{\left(  {g}%
_{14}-\frac{45}{8}\right)  }\right)  }{{\tilde{l}}^{2}}\nonumber\\
&  +{\tilde{S}}^{2}\frac{{g}_{10}^{b}\left(  3+e^{2}\right)  }{\tilde{l}^{3}%
}+\frac{45}{8}\frac{{\tilde{S}}^{4}\left(  3+e^{2}\right)  ^{2}}{\tilde{l}%
^{5}}\Bigr ]. \label{eq:dQdt-2PN-Barausse-body}%
\end{align}
But it is preferable that the formula for $dQ/dt$ (and for $d\iota/dt$) that
we test be accompanied, in the same work, by their associated expressions for
$d{\tilde{l}}/dt$ and $d{e}/dt$; and fortunately a paper by Ganz et al.
\cite{2007PThPh.117.1041G} provides such information, which we shall use in
our analysis. In particular, we will use equation (4.3) in
\cite{2007PThPh.117.1041G}: the evolution equation for $\tilde{l}$,%
\begin{align}
\frac{d{\tilde{l}}}{dt}  &  =-\frac{64}{5}\left(  \frac{{m}}{{M}^{2}}\right)
{\tilde{l}}^{-3}{\left(  1-{e}^{2}\right)  ^{\frac{3}{2}}}\nonumber\\
&  \times\left[  {{g_{9}-\frac{f_{1}}{\tilde{l}}+\pi\frac{g_{12}}{\tilde
{l}^{3/2}}+\frac{f_{3}-{f}_{4}{{\tilde{S}^{2}}}}{\tilde{l}^{2}}-\pi}}%
\frac{f_{7}}{\tilde{l}^{5/2}}\right. \nonumber\\
&  \left.  {{+\frac{f_{2}{{\left(  3+e^{2}\right)  \tilde{S}^{2}}}}{\tilde
{l}^{3}}-}}\frac{f_{6}{{\left(  3+e^{2}\right)  \tilde{S}^{2}}}}{\tilde{l}%
^{4}}{{+\frac{f_{5}{{\left(  3+e^{2}\right)  }}^{2}{{\tilde{S}^{4}}}}%
{\tilde{l}^{5}}}}\right]  , \label{eq:dldt-Ganz-body}%
\end{align}
which is (excluding the common factor, $\tilde{l}^{-3}$) to $O\left(
{\tilde{l}}^{-5/2}\right)  $ in \cite{2007PThPh.117.1041G}; and the evolution
equation for $e$,%
\begin{align}
\frac{de}{dt}  &  =-\frac{304}{15}\left(  \frac{{m}}{{M}^{2}}\right)
{\tilde{l}}^{-4}{\left(  1-{e}^{2}\right)  ^{\frac{3}{2}}}\nonumber\\
&  \times\left[  h_{1}-\frac{h_{2}}{\tilde{l}}{{+\pi\frac{h_{4}}{\tilde
{l}^{3/2}}-\frac{h_{5}+{h}_{6}{{\tilde{S}^{2}}}}{\tilde{l}^{2}}-\pi}}%
\frac{h_{9}}{\tilde{l}^{5/2}}\right. \nonumber\\
&  \left.  {{+\frac{h_{3}{{\left(  3+e^{2}\right)  \tilde{S}^{2}}}}{\tilde
{l}^{3}}-}}\frac{h_{8}{{\left(  3+e^{2}\right)  \tilde{S}^{2}}}}{\tilde{l}%
^{4}}{{+\frac{h_{7}{{\left(  3+e^{2}\right)  }}^{2}{{\tilde{S}^{4}}}}%
{\tilde{l}^{5}}}}\right]  , \label{eq:dedt-Ganz-body}%
\end{align}
which is (excluding the common factor, $\tilde{l}^{-4}$) also to $O\left(
{\tilde{l}}^{-5/2}\right)  $ in \cite{2007PThPh.117.1041G}.

The evolution equation of $Q$,
\begin{align}
\left(  \frac{d{Q}}{d{t}}\right)  _{2.5PN}  &  =-\frac{64}{5}\left(  \frac
{{m}}{{M}^{2}}\right)  {\tilde{l}}^{-3}{\left(  1-{e}^{2}\right)  ^{\frac
{3}{2}}}\left(  1-\frac{\tilde{S}^{2}\left(  3+e^{2}\right)  ^{2}}{\tilde
{l}^{3}}\right)  \times\left[  g_{9}\right. \nonumber\\
&  -\frac{d_{1}}{\tilde{l}^{1}}+\pi\frac{g_{12}}{\tilde{l}^{3/2}}-\frac
{d_{3}-d_{4}\tilde{S}^{2}}{\tilde{l}^{2}}-\pi\frac{d_{7}}{\tilde{l}^{5/2}%
}\nonumber\\
&  \left.  +\frac{\tilde{S}^{2}\left(  3+e^{2}\right)  d_{2}}{\tilde{l}^{3}%
}-\frac{\tilde{S}^{2}\left(  3+e^{2}\right)  d_{6}}{\tilde{l}^{4}}%
+\frac{\tilde{S}^{4}\left(  3+e^{2}\right)  ^{2}d_{5}}{\tilde{l}^{5}}\right]
, \label{eq:dQdt-Ganz-body}%
\end{align}
which corresponds to equation (4.1) in \cite{2007PThPh.117.1041G}, was of
$O\left(  {\tilde{l}}^{-5/2}\right)  $ (excluding the common factor,
$\tilde{l}^{-3}$). These formulae (equations (\ref{eq:dldt-Ganz-body}),
(\ref{eq:dedt-Ganz-body}), and (\ref{eq:dQdt-Ganz-body})) have been converted
from the variables used in \cite{2007PThPh.117.1041G} to our variables (see
Appendices \ref{2PN-flux-for-Q} and \ref{Evolution-Equaions-for-l-e-i}).
Because some of the original coefficients contained $\cos\left(  \iota\right)
$, which we have replaced with the approximation based on equation
(\ref{eq:LzQ_Series_First_Order_III}), there are new terms, which correspond
to $\tilde{l}^{-3}$, $\tilde{l}^{-4}$, and $\tilde{l}^{-5}$, in each of
equations (\ref{eq:dldt-Ganz-body}), (\ref{eq:dedt-Ganz-body}), and
(\ref{eq:dQdt-Ganz-body}). The original expressions did not include terms with
these powers of $\tilde{l}$, so we cannot use the new terms to extend the
accuracy of our analysis beyond that of the original expressions in Ganz et
al. \cite{2007PThPh.117.1041G}. Further, these evolution equations are
$O\left(  e^{2}\right)  $, hence the final results must also be used up to the
second power of $e$.%

\begin{table}[tbp] \centering
\caption[The coefficients of $-5/64\ M^{2}/m\ \tilde{l}^{3}{(1-e^{2})^{-3/2}}dQ/dt$ up to $\tilde{l}^{-3}$.]{The coefficients of $-5/64\ M^{2}/m\ \tilde{l}^{3}{(1-e^{2})^{-3/2}}dQ/dt$. \ The first column contains our calculated results at the
abutment. \ The second column contains the results of Ganz et al.
\protect\cite{2007PThPh.117.1041G}, and the third column contains the
results of Barausse, Hughes, and Rezzolla \protect\cite{2007PhRvD..76d4007B}. \ The trigonometric functions in both of these sets were evaluated on the
abutment.  Note: although the terms are reported to $O(e^{4})$ they are only accurate to $O(e^{2})$.}\label{Comparison-Table}%

\begin{tabular}
[c]{cccccc}
&  &  &  &  & \\\hline
& Results at the Abutment &  & (equation 4.1 in \cite{2007PThPh.117.1041G}) &
& (equation (A3) in \cite{2007PhRvD..76d4007B})\\
& (see equation (\ref{eq:Q_dot_III})) &  & (see equation
(\ref{eq:dQdt-Ganz-body})) &  & (see equation (\ref{eq:dQdt-2PN-Barausse-body}%
))\\\hline
\multicolumn{1}{|c}{$\tilde{l}^{0}$} & \multicolumn{1}{l}{$1+\frac{7}{8}e^{2}
$} &  & \multicolumn{1}{l}{$1+\frac{7}{8}e^{2}$} &  &
\multicolumn{1}{c|}{$1+\frac{7}{8}e^{2}$}\\
\multicolumn{1}{|c}{$\tilde{l}^{-1}$} & \multicolumn{1}{l}{$-\left(
\frac{743}{336}-\frac{23}{42}e^{2}-\frac{121}{96}e^{4}\right)  $} &  &
\multicolumn{1}{l}{$-\left(  \frac{743}{336}-\frac{23}{42}e^{2}\right)  $} &
& \multicolumn{1}{c|}{$-\left(  \frac{743}{336}-\frac{23}{42}e^{2}-\frac
{7}{16}e^{4}\right)  $}\\
\multicolumn{1}{|c}{$\tilde{l}^{-3/2}$} & \multicolumn{1}{l}{$\pi\left(
4+\frac{97}{8}e^{2}\right)  $} &  & \multicolumn{1}{l}{$\pi\left(  4+\frac
{97}{8}e^{2}\right)  $} &  & \multicolumn{1}{c|}{$\pi\left(  4+\frac{97}%
{8}e^{2}\right)  $}\\
\multicolumn{1}{|c}{$\tilde{l}^{-2}$} & \multicolumn{1}{l}{$-\left[
\frac{129293}{18144}+\frac{84035}{1728}e^{2}-\frac{629}{672}e^{4}\right.  $} &
& \multicolumn{1}{l}{$-\left[  \frac{129293}{18144}+\frac{84035}{1728}%
e^{2}\right.  $} &  & \multicolumn{1}{c|}{$-\left[  \frac{129293}{18144}%
+\frac{84035}{1728}e^{2}-\frac{575}{336}e^{4}\right.  $}\\
\multicolumn{1}{|c}{} & \multicolumn{1}{r}{$\left.  +\tilde{S}^{2}\left(
\frac{329}{96}+\frac{929}{96}e^{2}\right)  \right]  $} &  &
\multicolumn{1}{r}{$\left.  +\tilde{S}^{2}\left(  \frac{329}{96}+\frac
{929}{96}e^{2}\right)  \right]  $} &  & \multicolumn{1}{c|}{$\left.
+\tilde{S}^{2}\left(  \frac{342}{96}-\frac{570}{96}e^{2}\right)  \right] \S $%
}\\\cline{6-6}%
\multicolumn{1}{|c}{$\tilde{l}^{-5/2}$} & \multicolumn{1}{l}{$-\pi\left(
\frac{4159}{672}+\frac{21229}{1344}e^{2}-\frac{41783}{1344}e^{4}\right)  $} &
& \multicolumn{1}{l}{$-\pi\left(  \frac{4159}{672}+\frac{21229}{1344}%
e^{2}\right)  $} &  & \multicolumn{1}{|c}{$\pi\left(  \frac{4032}{672}%
+\frac{27132}{1344}e^{2}+\frac{8148}{1344}e^{4}\right)  $}\\\cline{1-5}%
\cline{2-5}%
$\tilde{l}^{-3}$ & \multicolumn{1}{l}{$-\left[  \frac{3819}{112}+\frac
{770993}{12096}e^{2}+\frac{333937}{2304}e^{4}\right.  $} &  &
\multicolumn{1}{l}{$-\left[  0\right.  $} &  & $-$\\
& \multicolumn{1}{r}{$\left.  -\tilde{S}^{2}\left(  \frac{250}{8}+\frac
{7261}{192}e^{2}-\frac{821}{384}e^{4}\right)   \right]  $} &  & \multicolumn{1}{r}{$\left.
-\tilde{S}^{2}\left(  \frac{183}{8}+\frac{607}{8}e^{2}+\frac{161}{8}%
e^{4}\right)  \right]  $} &  & $-$\\\hline
\end{tabular}
%

\end{table}%

We assume that the evolution of the orbit, $d\tilde{l}/dt$ and $de/dt$, is
described by equations (\ref{eq:dldt-Ganz-body}) and (\ref{eq:dedt-Ganz-body})
(from Ganz et al. \cite{2007PThPh.117.1041G}); the result of evaluating
equation (\ref{eq:Q_dot_III}) is compiled in table \ref{Comparison-Table}
(first column). The second column contains the result $\left(  dQ/dt\right)
_{2.5PN}$, derived by Ganz et al. (see equation (\ref{eq:dQdt-Ganz-body})),
evaluated on the abutment. Similarly, the third column contains the formula
for $\left(  dQ/dt\right)  _{2PN}$ based on equation (A.3) in
\cite{2007PhRvD..76d4007B}, also evaluated on the abutment.

Although the terms in table \ref{Comparison-Table} are $O\left(  e^{4}\right)
$ (N.B. Those given to $O\left(  e^{2}\right)  $ were found to have no terms
of higher order in $e^{2}$), any final result one might calculate must be of
$O\left(  e^{2}\right)  $. To that order, our calculated results at the
abutment agree with those of Ganz et al. \cite{2007PThPh.117.1041G} up to
$\tilde{l}^{-5/2}$. There is also agreement with Barausse, Hughes, and
Rezzolla \cite{2007PhRvD..76d4007B} up to $\tilde{l}^{-2}$ with the exception
of the coefficient for the $\tilde{S}^{2}$ term in the third column (marked
with $\S$), which differs from the other two results;\ the expanded equation in
Barausse, Hughes, and Rezzolla differed from that of Ganz for that order of
$\tilde{l}$.

There are two reasons for reporting these results to $O\left(  e^{4}\right)
$: first, we wish to demonstrate that the confirmation of Ganz's calculations
is not to be dismissed as a fortuitous triviality. The method of calculation
of $dQ/dt$ at the abutment differs fundamentally from that used by Ganz et al.
to derive their results, as would be required of a good consistency condition;
second, the differing coefficient values for the $e^{4}$ terms demonstrate
that one must not perform calculations on the abutment for highly eccentric
orbits. While the abutment equations ($\partial Q_{X}/\partial\tilde{l}$,
$\partial Q_{X}/\partial e$, and $\iota\left(  \tilde{l},e,\tilde{S}\right)
$) are exact in terms of $e^{2}$, one remains limited by the order of $e$ used
in the radiation back-reaction model being tested.

Because the expressions for $\partial Q_{X}/\partial\tilde{l}$ and $\partial
Q_{X}/\partial e$ can be derived to arbitrary order in $\tilde{l}^{-1}$, and
the coefficients for each power are exact finite series in $e^{2}$, it is
worthwhile to consider using the abutment to improve the order of $e^{2}$ of
the evolution equations in the weaker field regime by allowing other
theoreticians to perform a test of their own, improved back-reaction models.
Since the abutment extends down to the LSO, one might also explore the
development and testing of evolution equations in the strong-field regime,
given that on the abutment the trigonometric contributions of $\sin\left(
\iota\right)  $ and $\cos\left(  \iota\right)  $ can be expressed as functions
of $e$, $\tilde{l}$, and $\tilde{S}$. But one must also be mindful of the
assumptions made at the outset of this exercise, in particular, the assumption
that the secondary object can be approximated as a test-particle of
infinitesimal mass, and the use of adiabatically evolving orbits.

\subsection{\label{Second-Order-Calculation-of-diota-by-dt}The Second-order
Calculation of $d\iota/dt$\ for the Leading Order of $\tilde{S}$ (weak-field
regime)}

Now that $\wp(e,\tilde{l})$ is known we can calculate $g\left(  e,\tilde
{l}\right)  $; but let us first derive $de/d\tilde{l}$ using equations
(\ref{eq:dldt-Ganz-body}) and (\ref{eq:dedt-Ganz-body}). We find the
following:%
\begin{equation}
\frac{de}{d\tilde{l}}=\frac{19}{12}e\left(  \frac{1-\frac{145}{304}e^{2}%
}{\tilde{l}}+\frac{\frac{3215}{3192}-\frac{33373}{102144}e^{2}}{\tilde{l}^{2}%
}\right)  ,
\end{equation}
expressed to $O\left(  {\tilde{l}}^{-2}\right)  $. From equation
(\ref{eq:Reductive-Ansatz-Elliptical}), one derives:%
\begin{equation}
g\left(  e,\tilde{l}\right)  =\tilde{S}^{2}\left[  \frac{a_{0}\left(
e\right)  }{\tilde{l}^{4}}+\frac{a_{1}\left(  e\right)  }{\tilde{l}^{5}%
}\right]  ,
\end{equation}
where%
\[
a_{0}\left(  e\right)  =a_{0}\left(  0\right)  \left(  1-\frac{119}{432}%
b_{0}e^{2}\right)
\]
and%
\[
a_{1}\left(  e\right)  =a_{1}\left(  0\right)  \left(  1+\left(  \frac
{3215}{72576}b_{0}-\frac{143}{864}b_{1}\right)  e^{2}\right)  ,
\]
which can be used to calculate $\partial{\iota}/\partial{\tilde{l}}$ under the
reductive ansatz elliptical,%
\begin{align}
\frac{\partial\iota}{\partial\tilde{l}}  &  =-\tilde{S}\left\{  \frac{5}%
{2}\left(  3+e^{2}\right)  \left(  1+e^{2}\right)  +\frac{1}{2}\left(
3+e^{2}\right)  A_{0}\left(  e\right)  -\frac{1}{4}A_{1}\left(  e\right)
\right\}  \tilde{l}^{-7/2}\nonumber\\
&  -\tilde{S}\left\{  \frac{3}{2}\left(  3+e^{2}\right)  -A_{0}\left(
e\right)  \right\}  \tilde{l}^{-5/2}, \label{eq:didl_Reduced_Elliptical}%
\end{align}
where%
\[
A_{0}\left(  e\right)  =A_{0}\left(  0\right)  \left(  1-\frac{1}{2}\frac
{119}{432}b_{0}e^{2}\right)
\]
and%
\[
A_{1}\left(  e\right)  =\frac{a_{1}\left(  e\right)  }{A_{0}\left(  e\right)
}.
\]

Now that we have developed a formula for $\partial\iota/\partial\tilde{l}$
(equation (\ref{eq:didl_Reduced_Elliptical})) that incorporates the reductive
ansatz elliptical, and we have found that $\partial\iota/\partial e$
(equation(\ref{eq:dide})) is unaffected by the reductive ansatz elliptical,
the expression for $d\iota/dt$ can be obtained from equation
(\ref{eq:full_by_t}) to the leading order in $\tilde{S}$ with coefficients of
$O\left(  e^{2}\right)  $:%
\begin{equation}
\frac{d\iota}{dt}=\tilde{S}\frac{m}{M^{2}}\tilde{l}^{-4}{\left(  1-{e}%
^{2}\right)  ^{\frac{3}{2}}}\left(  \frac{U_{1}}{\tilde{l}^{3/2}}-\frac{U_{3}%
}{\tilde{l}^{5/2}}\right)  , \label{eq:iota-dot-reductive}%
\end{equation}
where%
\begin{align*}
U_{1}  &  =\frac{32}{15}\left(  9-2A_{0}\left(  e\right)  \right)  +\frac
{4}{15}\left(  -109+42A_{0}(e)\right)  e^{2}\\
U_{3}  &  =\frac{2}{105}\left(  1647-2494A_{0}\left(  e\right)  +168A_{1}%
\left(  e\right)  \right) \\
&  +\frac{1}{105}\left(  682-2978A_{0}\left(  e\right)  +147A_{1}\left(
e\right)  \right)  e^{2}.
\end{align*}

Equation (\ref{eq:didt_Ganz_Appendix}) can be expanded and expressed to
leading order in $\tilde{S}$ to yield:%
\begin{equation}
\frac{d\iota}{dt}=\frac{244}{15}\tilde{S}\frac{m}{M^{2}}\tilde{l}^{-4}{\left(
1-{e}^{2}\right)  ^{\frac{3}{2}}}\left(  \frac{u_{1}}{\tilde{l}^{3/2}}%
-\frac{u_{3}}{\tilde{l}^{5/2}}\right)  . \label{eq:didt-Ganz-Body}%
\end{equation}
By equating the terms in equations (\ref{eq:iota-dot-reductive}) and
(\ref{eq:didt-Ganz-Body}) (i.e. $U_{1}=244u_{1}/15$ and $U_{3}=244u_{3}/15$)
one first solves for $A_{0}\left(  0\right)  $ and $A_{1}\left(  0\right)  $
for a circular orbit by setting $e=0$:%
\[
A_{0}\left(  0\right)  =\frac{155}{48}%
\]
and%
\[
A_{1}\left(  0\right)  =\frac{279289}{4032}.
\]
By substituting these values into equation (\ref{eq:iota-dot-reductive}), we
obtain,%
\begin{align}
\frac{d\iota}{dt}  &  =\frac{244}{15}\tilde{S}\frac{m}{M^{2}}\tilde{l}%
^{-4}{\left(  1-{e}^{2}\right)  ^{\frac{3}{2}}}\nonumber\\
&  \times\left[  \frac{\left(  1+\left(  \frac{18445}{52704}b_{0}-\frac
{213}{488}\right)  e^{2}\right)  }{\tilde{l}^{3/2}}\right. \nonumber\\
&  \left.  -\frac{\left(  \frac{10461}{1708}+\left(  \frac{79869}{54656}%
-\frac{39938327}{17708544}b_{1}+\frac{5621763839}{1487517696}b_{0}\right)
e^{2}\right)  }{\tilde{l}^{5/2}}\right]  . \label{eq:didt-Result}%
\end{align}
When evaluated at $e=0$, the expression in equation (\ref{eq:didt-Result})
matches the results reported in the literature (equations
(\ref{eq:iota-dot-Ryan}), (\ref{eq:iota-dot-Flanagan-Hinderer}), and
(\ref{eq:didt-Ganz-Body})) for near-polar orbits. For near-circular orbits,
values of $b_{0}$ and $b_{1}$ can be found for which the coefficients of the
\ $\tilde{l}^{-3/2}$ and $\tilde{l}^{-5/2}$\ terms in equation
(\ref{eq:didt-Result}) match their theoretical counterparts in equation
(\ref{eq:didt-Ganz-Body}).

\subsection{The Independence of the Abutment of Radiation Back-reaction
Models}

Let us clarify the meaning of our statement that the abutment model is
independent of any specific radiation back-reaction model. The expression for
the abutment, $Q_{X}$ (equation (\ref{eq:Q_Abutment_Analytical})), is
determined by the characteristics of the Kerr spacetime of the primary object
in which the secondary object (i.e. test-particle) orbits. The analytical
expressions for $d\tilde{l}/dt$ and $de/dt$ describe the effects of radiation
back-reaction on the values of $\tilde{l}$ and $e$ of the orbit, and they
serve as inputs to our abutment model in two ways: first, through the quotient
$\epsilon/\lambda\cong\partial e/\partial\tilde{l}$ (equation
(\ref{eq:Reductive-Ansatz-Elliptical})); and second, through the directional
derivatives in equations (\ref{eq:Q_dot_III}) and (\ref{eq:full_by_t}).

The mechanics of the abutment remain consistent, the details of the radiation
back-reaction model notwithstanding. The results of either directional
derivative are outputs of the abutment model that describe the effect of the
radiation back-reaction on the listing of the test-particle orbit.

\section{\label{Conclusions}Conclusions}

For inclined test-particle orbits around a black hole, two solutions for
${X}^{2}$ (where $X=\tilde{L}_{z}-\tilde{S}\tilde{E}$) can be derived: ${X}_{-}^{2}$
and ${X}_{+}^{2}$. Given a Schwarzschild black hole (SBH), ${X}_{+}^{2}%
={X}_{-}^{2}$ on any polar orbit, where ${X}_{-}^{2}$ corresponds to prograde
orbits and ${X}_{+}^{2}$ corresponds to retrograde orbits. For a Kerr black
hole (KBH) the orbits on which ${X}_{+}^{2}={X}_{-}^{2}$ are not polar, but
near-polar and retrograde. Such orbits comprise the abutment at which the
value of the Carter constant ($Q$) is a maximum for given values of latus
rectum ($\tilde{l}$) and eccentricity ($e$).

In this work we derived an analytical formula for the value of orbital
inclination, $\iota$, of an elliptical orbit on the abutment. By performing
the partial differentiation of $\iota$ with respect to $\tilde{l}$, we were
able to confirm the numerical result for $\partial\iota/\partial\tilde{l}$
reported in Komorowski et al. \cite{Komorowski:2010we} \ for circular orbits,
and we were able to extend the formula to include $\partial\iota
/\partial\tilde{l}$ for elliptical orbits. A result for $\partial
\iota/\partial e$ was also obtained for elliptical orbits. Further, it allowed
one to redefine, in terms of $e$, $\tilde{l}$, and $\tilde{S}$, any
trigonometric function that might be found in an evolution equation to be
tested at the abutment.

Evolving orbits in Kerr spacetime are not constrained to follow the abutment.
Instead, the value of $Q$ will follow $Q_{path}$, which intersects the
abutment tangentially at an arbitrary point of contact of the first order.
This behaviour is assured because the value of $Q_{path}$ cannot exceed that
of $Q$ on the abutment; to do so would make ${X}_{\pm}^{2}$ complex and thus
unphysical. For circular orbits, we modelled the second-order behaviour
reported in \cite{Komorowski:2010we} by introducing a bounded function
$f\left(  e,\tilde{l}\right)  $ (also in terms of $\tilde{S}$) to reduce the
value of $\partial^{2}Q_{path}/\partial\tilde{l}^{2}$ while leaving $Q_{path}$
and $\partial Q_{path}/\partial\tilde{l}$ equal to their corresponding values
($Q_{X}$ and $\partial Q_{X}/\partial\tilde{l}$) on the abutment. This
approach was then applied to elliptical orbits, and a new bounded function
$g\left(  e,\tilde{l}\right)  $, which depends upon $de/d\tilde{l}$, was used
to reduce the value of $\partial^{2}Q_{X}/\partial\tilde{l}^{2}$ to
$\partial^{2}Q_{path}/\partial\tilde{l}^{2}$. It was discovered that the value
of $\partial^{2}Q_{path}/\partial e^{2}$ remained unchanged by the reductive
ansatz elliptical.

The consistency of published evolution equations, $dQ/dt$, $d\tilde{l}/dt$,
and $de/dt$, was tested by using $d\tilde{l}/dt$ and $de/dt$ to generate an
expression for $dQ/dt$ at the abutment. In general, the calculation of $dQ/dt$
is more difficult to perform than that of $d\tilde{l}/dt$ and $de/dt$
\cite{PhysRevD.55.3444}; hence, the abutment provides a useful mechanism for
testing the validity of radiation back-reaction models. Indeed, the evolution
equations reported by Ganz et al. \cite{2007PThPh.117.1041G} were confirmed to
their 2.5PN order. The abutment provides a consistency condition that is
limited to near-polar retrograde orbits; and yet, some back-reaction models
have been found to exhibit pathological behaviour for polar orbits (Gair and
Glampedakis \cite{2006PhRvD..73f4037G}). Another consistency condition is
already known for orbits of any $\iota$ (see equations (13) and (14) in
\cite{2006PhRvD..73f4037G}), but it applies to circular orbits (i.e. in the
limit $e\rightarrow0$); the abutment is valid for orbits of arbitrary
eccentricity, depending on the accuracy of the back-reaction model used.

This method promises to be a useful tool for confirming the accuracy of
evolution equations to greater order in $e$ and $\tilde{l}^{-1}$. Further work
might entail the development of a more precise mathematical treatment of the
ans\"{a}tze in relation to the underlying physical concepts of the radiation
back-reaction process and its effect on the listing behaviour of orbits near
the abutment. It would also be intriguing to investigate the consistency
condition reported by Gair and Glampedakis \cite{2006PhRvD..73f4037G}, on the abutment.

\begin{acknowledgments}
The authors are very grateful to Dr. Daniel Kennefick for his continued
encouragement and for his invaluable discussions of this manuscript. MH's
research is funded through the NSERC Discovery Grant, Canada Research Chair,
Canada Foundation for Innovation, Ontario Innovation Trust and Western's
Academic Development Fund programs. SRV would like to acknowledge the Faculty
of Science for a UWO Internal Science Research Grant award during the progress
of this work.
\end{acknowledgments}

\newpage

\appendix

\section{Ancillary Equations}%

\setcounter{equation}{0}
\renewcommand{\theequation}{\Alph{section}\arabic{equation}}%

\subsection{\label{Taylors_Series_for_two_Variables}Taylor Series for two
Variables}

Refer to Chapter 6 in \cite{Spiegel:1963fk} for a more detailed treatment. Let
us consider a locally continuous function with two independent variables,
$f(x,y)$. We may use an operator%

\begin{equation}
\left(  h\frac{\partial}{\partial{x}}+k\frac{\partial}{\partial{y}}\right)
\end{equation}
to construct a Taylor series of $n$ terms%
\begin{align}
f\left(  x_{o}+h,y_{o}+k\right)   &  =f\left(  x_{o},y_{o}\right) \nonumber\\
&  +\left[  \left(  h\frac{\partial}{\partial{x}}+k\frac{\partial}{\partial
{y}}\right)  f\left(  x,y\right)  \right]  _{{x}=x_{o},{{y}=y_{o}}}+\frac
{1}{2!}\left[  \left(  h\frac{\partial}{\partial{x}}+k\frac{\partial}%
{\partial{y}}\right)  ^{2}f\left(  x,y\right)  \right]  _{{x}=x_{o},{{y}%
=y_{o}}}\nonumber\\
&  \ldots+\frac{1}{n!}\left[  \left(  h\frac{\partial}{\partial{x}}%
+k\frac{\partial}{\partial{y}}\right)  ^{n}f\left(  x,y\right)  \right]
_{{x}=x_{o},{{y}=y_{o}}}%
\end{align}
if the $(n+1)^{th}$ partial derivatives are continuous. In this paper, we are
concerned only with the second derivative.

\subsection{\label{Generalised_Treatment_of_the_Taylors_Series}Treatment of
the Taylor Series Under Partial Differentiation}

Given the term:
\begin{equation}
A=hf\left(  x_{o}\right)  g\left(  x\right)  ,
\end{equation}
where $h=\left(  x-x_{o}\right)  $. We can calculate the partial derivative of
$A$ with respect to $x$,
\begin{align}
\frac{\partial A}{\partial x}  &  =\frac{\partial}{\partial x}\left(
hf\left(  x_{o}\right)  g\left(  x\right)  \right) \nonumber\\
&  =f\left(  x_{o}\right)  \frac{\partial}{\partial x}\left(  hg\left(
x\right)  \right) \nonumber\\
&  =f\left(  x_{o}\right)  \left(  h\frac{\partial}{\partial x}g\left(
x\right)  +g\left(  x\right)  \frac{\partial}{\partial x}h\right)  ,
\end{align}
and thus demonstrate%

\begin{equation}
\left.  \frac{\partial A}{\partial x}\right\vert _{x=x_{o}}=f\left(
x_{o}\right)  g\left(  x_{o}\right)  .
\end{equation}

Consider a more complicated case where we have a function $F(x,h,y,k)$, where
$h=\left(  x-x_{o}\right)  $ and $k=\left(  y-y_{o}\right)  $. Calculate
\begin{equation}
\left.  \frac{\partial F}{\partial x}\right\vert _{x=x_{o}}\text{and}\left.
\frac{\partial F}{\partial y}\right\vert _{y=y_{o}}.
\end{equation}
If we hold $y$ constant and set $k=0$, then%

\begin{equation}
\frac{dF}{dx}=\frac{\partial F}{\partial x}+\frac{\partial F}{\partial h}%
\frac{\partial h}{\partial x}. \label{Total_Derivative_dFdx}%
\end{equation}
If we hold $x$ constant and set $h=0$, then%

\begin{equation}
\frac{dF}{dy}=\frac{\partial F}{\partial y}+\frac{\partial F}{\partial k}%
\frac{\partial k}{\partial y}. \label{Total_Derivative_dFdy}%
\end{equation}
These results will be of use in applying the second-order effects to
$Q_{path}$ as it makes contact with the abutment, $Q_{X}$.

\subsection{\label{Treatment-of-Qx-series}Treatment of $Q_{X}$\ as a Series in
$\tilde{l}$}

The expansion of $Q_{X}$ in terms of $\tilde{S}$\ (equation
(\ref{eq:Q_Abutment_Intermediate_Form_2_III})) can be expressed as a series in
$\tilde{l}$:%
\begin{equation}
Q_{X}=\tilde{l}+\sum_{i=0}^{\infty}\left[  {\frac{\left(  3+{e}^{2}\right)
^{i+1}}{{\tilde{l}}^{i}}+\frac{i\left(  i-1\right)  }{2}}\frac{{\left(
3+{e}^{2}\right)  ^{i-2}\left(  1-e^{2}\right)  }^2}{{\tilde{l}}^{i}}{\tilde
{S}^{2}}\right]  . \label{eq:Qx-l-series}%
\end{equation}

From equation (\ref{eq:Qx-l-series}) one can obtain $\partial Q_{X}%
/\partial\tilde{l}$ and $\partial Q_{X}/\partial e$ directly:%
\begin{equation}
\frac{\partial Q_{X}}{\partial\tilde{l}}=1-\sum_{i=0}^{\infty}\left[
{\frac{i\left(  3+{e}^{2}\right)  ^{i+1}}{{\tilde{l}}^{i+1}}+\frac
{i^{2}\left(  i-1\right)  }{2}}\frac{{\left(  3+{e}^{2}\right)  ^{i-2}\left(
1-e^{2}\right)  }^2}{{\tilde{l}}^{i+1}}{\tilde{S}^{2}}\right]  ,
\label{eq:dQxdl-appendix}%
\end{equation}
and%
\begin{align}
\frac{\partial Q_{X}}{\partial e}  &  =2e\sum_{i=0}^{\infty}\left[
{\frac{\left(  i+1\right)  \left(  3+{e}^{2}\right)  ^{i}}{{\tilde{l}}^{i}%
}-{{i\left(  i-1\right)  }}\frac{{\left(  3+{e}^{2}\right)  ^{i-2}\left(  1-{e}^{2}\right) }%
}{{\tilde{l}}^{i}}\tilde{S}^{2}}\right. \nonumber\\
&  \left.  {+\frac{i\left(  i-1\right)  \left(  i-2\right)  }{2}}%
\frac{{\left(  3+{e}^{2}\right)  ^{i-3}\left(  1-e^{2}\right)^2  }}{{\tilde{l}%
}^{i}}{\tilde{S}^{2}}\right]  . \label{eq:dQxde-appendix}%
\end{align}
The $\sqrt{Q_{X}}$ will also be required for the treatment of $dQ/dt$, (see
Appendix \ref{2PN-flux-for-Q})%
\begin{align}
\sqrt{Q_{X}}  &  =\sqrt{{\tilde{l}}}+\frac{1}{2}{\frac{\left(  3+{e}%
^{2}\right)  }{\sqrt{{\tilde{l}}}}}+\frac{3}{8}\,{\frac{\left(  3+{e}%
^{2}\right)  ^{2}}{{\tilde{l}}^{3/2}}}+\frac{1}{2}\frac{\left(  \left(
1-e^{2}\right)^2  {\tilde{S}}^{2}+{\frac{5}{8}}\,\left(  3+{e}^{2}\right)
^{3}\right)  }{{\tilde{l}}^{5/2}}\nonumber\\
&  +{\frac{5}{128}}\,{\frac{\left(  3+{e}^{2}\right)  \left(  32\,\left(
1-e^{2}\right)^2  {\tilde{S}}^{2}+7\,\left(  3+{e}^{2}\right)  ^{3}\right)
}{{\tilde{l}}^{7/2}}}\nonumber\\
&  +{\frac{7}{256}}\,{\frac{\left(  3+{e}^{2}\right)  ^{2}\left(  80\,\left(
1-e^{2}\right)^2  {\tilde{S}}^{2}+9\,\left(  3+{e}^{2}\right)  ^{3}\right)
}{{\tilde{l}}^{9/2}}+\ldots} \label{eq:QX-Square-Root}%
\end{align}

\subsection{\label{2PN-flux-for-Q}The 2PN Flux for $Q$}

Equation (\ref{eq:dQdt-2PN-Barausse-body}), an expression for $dQ/dt$, was
derived from equation (A.3) in \cite{2007PhRvD..76d4007B} (after equation (56)
in \cite{2006PhRvD..73f4037G}) by substituting the approximations of
$\sin\left(  \iota\right)  $ and $\cos\left(  \iota\right)  $ on the abutment.
where%
\[
{g}_{9}=1+\frac{7}{8}{e}^{2},\quad{g}_{10}^{b}=\frac{61}{8}+\frac{91}{4}%
{e}^{2}+\frac{461}{64}{e}^{4},\quad{g}_{11}=\frac{1247}{336}+\frac{425}%
{336}{e}^{2},
\]%
\[
{g}_{12}=4+\frac{97}{8}{e}^{2},\quad{g}_{13}=\frac{44711}{9072}+\frac
{302893}{6048}{e}^{2},\quad{g}_{14}=\frac{33}{16}+\frac{95}{16}{e}^{2}.
\]

An alternative expression for $d{Q}/d{t}$ to 2.5PN order was presented by Ganz
et al. (equation (4.1) in \cite{2007PThPh.117.1041G}) in which we have
converted their variable, $Y\simeq\cos\left(  \iota\right)  $, to $-\tilde
{S}\left(  3+e^{2}\right)  \tilde{l}^{-3/2}$ to yield equation
(\ref{eq:dQdt-Ganz-body}), where%
\[
{d}_{1}=\frac{743}{336}-\frac{23}{42}e^{2},\quad d_{2}=\frac{85}{8}+\frac
{211}{8}e^{2}\quad{d}_{3}=\frac{129193}{18144}+\frac{84035}{1728}e^{2},
\]%
\[
{d}_{4}=\frac{329}{96}+\frac{929}{96}e^{2},\quad{d}_{5}=\frac{53}{8}%
+\frac{163}{8}e^{2},\quad{d}_{6}=\frac{2553}{224}-\frac{553}{192}e^{2},\quad
d_{7}=\frac{4159}{672}+\frac{21229}{1344}e^{2}.
\]

\subsection{\label{Evolution-Equaions-for-l-e-i}Evolution Equations for
$\tilde{l}$, $e$, and $\iota$}

The evolution equations for $\tilde{l}$, $e$ , and $\iota$ are reported by
Ganz et al. (see equation (4.3) in \cite{2007PThPh.117.1041G}) to $O\left(
\tilde{l}^{-5/2}\right)  $. We reproduce them in equations
(\ref{eq:dldt-Ganz-body}), (\ref{eq:dedt-Ganz-body}), and
(\ref{eq:didt_Ganz_Appendix}) after having converted their original variable,
$\upsilon=\sqrt{\frac{M}{l}}$, to $\tilde{l}^{-1/2}$, and by using
$d\upsilon=-1/2\tilde{l}^{-3/2}d\tilde{l}$, where%
\[
f_{1}=\frac{743}{336}+\frac{55}{21}e^{2},\quad f_{2}=\frac{133}{12}+\frac
{379}{24}e^{2},\quad f_{3}=\frac{34103}{18144}-\frac{526955}{12096}e^{2},
\]%
\[
f_{4}=\frac{329}{96}+\frac{929}{96}e^{2},\quad f_{5}=\frac{815}{96}+\frac
{477}{32}e^{2},\quad f_{6}=\frac{1451}{56}+\frac{1043}{96}e^{2},
\]%
\[
f_{7}=\frac{4159}{672}+\frac{48809}{1344}e^{2};
\]
and where%
\[
h_{1}=1+\frac{121}{304}e^{2},\quad h_{2}=\frac{6849}{2128}+\frac{4509}%
{2128}e^{2},\quad h_{3}=\frac{879}{76}+\frac{515}{76}e^{2},
\]%
\[
h_{4}=\frac{985}{152}+\frac{5969}{608}e^{2},\quad h_{5}=\frac{286397}%
{38304}+\frac{2064415}{51072}e^{2},\quad h_{6}=\frac{3179}{608}+\frac
{8925}{1216}e^{2},
\]%
\[
h_{7}=\frac{5869}{608}+\frac{10747}{1216}e^{2},{\quad}h_{8}=\frac{1903}%
{304}-\frac{22373}{8512}e^{2},{\quad}h_{9}=\frac{87947}{4256}+\frac
{4072433}{68096}e^{2}.
\]
We use the following equations to convert the form of the equation for list
rate in \cite{2007PThPh.117.1041G}:%
\begin{equation}
1-Y^{2}\simeq\sin^{2}\left(  \iota\right)  ;
\end{equation}%
\begin{align}
\frac{dY}{dt}  &  =\frac{d\cos\left(  \iota\right)  }{d\iota}\times
\frac{d\iota}{dt}\nonumber\\
&  =-\sin\left(  \iota\right)  \times\frac{d\iota}{dt}.
\label{eq:dYdi-appendix}%
\end{align}
We obtain:%
\begin{align}
\frac{d\iota}{dt}  &  =\frac{244}{15}\left(  \frac{{m}}{{M}^{2}}\right)
{\tilde{l}}^{-4}{\left(  1-{e}^{2}\right)  ^{\frac{3}{2}}}\left(  1-\frac
{1}{2}\tilde{S}^{2}\left(  3+e^{2}\right)  ^{2}\tilde{l}^{-3}\right)
\nonumber\\
&  \times\left[  \frac{u_{1}\tilde{S}}{\tilde{l}^{3/2}}-\frac{u_{3}\tilde{S}%
}{\tilde{l}^{5/2}}+\frac{u_{2}{{\left(  3+e^{2}\right)  }}\tilde{S}^{3}%
}{\tilde{l}^{7/2}}\right] \nonumber\\
&  =\frac{244}{15}\left(  \frac{{m}}{{M}^{2}}\right)  {\tilde{l}}^{-4}{\left(
1-{e}^{2}\right)  ^{\frac{3}{2}}}\nonumber\\
&  \times\left[  \frac{u_{1}\tilde{S}}{\tilde{l}^{3/2}}-\frac{u_{3}\tilde{S}%
}{\tilde{l}^{5/2}}+\frac{u_{2}{{\left(  3+e^{2}\right)  }}\tilde{S}^{3}%
}{\tilde{l}^{7/2}}\right. \nonumber\\
&  \left.  -\frac{1}{2}\frac{u_{1}\left(  3+e^{2}\right)  ^{2}\tilde{S}^{3}%
}{\tilde{l}^{9/2}}+\frac{1}{2}\frac{u_{3}\left(  3+e^{2}\right)  ^{2}\tilde
{S}^{3}}{\tilde{l}^{11/2}}-\frac{1}{2}\frac{u_{2}{{\left(  3+e^{2}\right)  }%
}^{3}\tilde{S}^{5}}{\tilde{l}^{13/2}}\right]  \label{eq:didt_Ganz_Appendix}%
\end{align}
where%
\[
u_{1}=1+\frac{189}{61}e^{2},\quad u_{2}=\frac{13}{244}+\frac{277}{244}%
e^{2},\quad u_{3}=\frac{10461}{1708}+\frac{83723}{3416}e^{2}.
\]

\section{\label{Apercu}Series Expansions of Critical Values in Terms of
$\tilde{S}$ and $\tilde{l}$}%

\setcounter{equation}{0}%

Our conversion of the quantities $X$ (equation (\ref{eq:X_S_Series_III})),
$\tilde{E}$ (equation (\ref{eq:E4_III})), and $Q_{X}$ (equation
(\ref{eq:Q_Abutment_Intermediate_Form_2_III})) to expansion series in
$\tilde{S}$ helped to simplify our analysis by avoiding the use of the much
more complicated series expansions in terms of $\tilde{l}$. Equation
(\ref{eq:cos-iota-third-order}) can be converted to a series:%
\begin{equation}
\overset{(2n+1)}{\frac{\tilde{L}_{z}}{\sqrt{Q_{X}}}}=\sum_{i=0}^{n}%
c_{2i+1}\tilde{S}^{2i+1}. \label{eq:c-S-series-appendix}%
\end{equation}
By choosing the order of $\tilde{S}$ (the value of $2n+1$) in which to work,
it becomes easier to derive suitable series approximations of these
quantities, and their mathematical combinations, in terms of $\tilde{l}$.
Since the equations derived during the full analytical treatment are
Brobdingnagian, and thus preclude detailed presentation in this paper, we
shall offer the essential highlights of our analysis.

\subsection{\label{First_Order_Calculations}First-order Calculations}

We require the series expansion of the quotient, which appears in equation
(\ref{eq:iota_first_order_in_S}),%
\begin{equation}
\overset{\left(  1\right)  }{\frac{{\tilde{L}}_{z}}{\sqrt{{Q}_{X}}}},
\end{equation}
to be expressed in terms of $\tilde{l}$. To obtain this result we perform a
careful manipulation of ${\tilde{L}}_{z}$ (in terms of $X$ and $\tilde{E}$,
\textit{viz.} equation (\ref{eq:calc_Lz})) and $Q_{X}$ (as a series expansion
in $\tilde{S}$) using MacLaurin series. The coefficient of $\tilde{S}^{1}$
(i.e. $c_{1}$) is converted to an expansion in $\tilde{l}$; we find $c_{1}$ to
be:%
\begin{equation}
c_{1}={\frac{\psi_{{1}}-\psi_{{2}}}{\sqrt{{\frac{{\tilde{l}}^{2}}{{\tilde{l}%
}-{e}^{2}-3}}}}} \label{eq:c1-appendix}%
\end{equation}
where%
\[
\psi_{{1}}=\sqrt{{\frac{{\tilde{l}}\left(  {\tilde{l}}^{2}-4\left(
\,{\tilde{l}}+\,{e}^{2}-1\right)  \right)  }{\left(  {\tilde{l}}-{e}%
^{2}-3\right)  ^{3}}}}%
\]%
\[
\psi_{{2}}=\sqrt{{\frac{{\tilde{l}}^{2}-4\,\left(  {\tilde{l}}+\,{e}%
^{2}-1\right)  }{{\tilde{l}}\left(  {\tilde{l}}-{e}^{2}-3\right)  }}}.
\]
From equation (\ref{eq:c1-appendix}), one obtains the result:%
\begin{equation}
c_{1}=-\left(  {e}^{2}+3\right)  \left(  {\frac{1}{{\tilde{l}}^{3/2}}}%
+{\frac{\left(  1+{e}^{2}\right)  }{{\tilde{l}}^{5/2}}}+{\frac{\left(
3+2\,{e}^{2}+{e}^{4}\right)  }{{\tilde{l}}^{7/2}}}+{\frac{\left(  9+5\,{e}%
^{2}+5\,{e}^{4}+{e}^{6}\right)  }{{\tilde{l}}^{9/2}}}\right)  ,
\end{equation}
which appears in equation (\ref{eq:LzQ_Series_First_Order_III}).

\subsection{\label{Third_Order_Calculations}Third-order\ Calculations}

The third-order calculations require two additional factors:%
\begin{equation}
\overset{\left(  3\right)  }{\frac{{\tilde{L}}_{z}}{\sqrt{{Q}_{X}}}%
\ }\text{and}\overset{\left(  2\right)  }{\ \frac{\tilde{S}^{2}\left(
{1-\tilde{E}^{2}}\right)  }{{Q}_{X}}}, \label{eq:factors-third-order-appendix}%
\end{equation}
which are used to evaluate $x=\cos(\iota)$ using equation
(\ref{eq:x-cos-iota-second-order}). The first factor can be derived by
converting the coefficient of $\tilde{S}^{3}$,%

\begin{equation}
c_{3}=-\frac{1}{2}\,{\frac{\left(  1-{e}^{2}\right)  ^{2}\left(  \left(
{\tilde{l}}^{2}-2\,{\tilde{l}}+2\,{\tilde{l}e}^{2}-16\,{e}^{2}\right)
\psi_{{1}}+\left(  -4+{\tilde{l}}\right)  \left(  {\tilde{l}}-2-2\,{e}%
^{2}\right)  \psi_{{2}}\right)  }{\sqrt{{\frac{{\tilde{l}}^{2}}{{\tilde{l}%
}-{e}^{2}-3}}}{\tilde{l}}\left(  {\tilde{l}}-{e}^{2}-3\right)  \left(
{\tilde{l}}^{2}-4\,{\tilde{l}}+4\,{e}^{2}-4\right)  },}%
\end{equation}
in equation (\ref{eq:c-S-series-appendix}), to a series expansion in
$\tilde{l}$ (see equation (\ref{eq:LzQ_Series_Third_Order_III})) and adding
the result to the first order term:
\begin{equation}
c_{3}=-{\tilde{S}}^{3}\left(  1-{e}^{2}\right)  ^{2}\left(  \frac{1}%
{{\tilde{l}}^{7/2}}+\frac{1}{2}\,{\frac{11+5\,{e}^{2}}{{\tilde{l}}^{9/2}}%
}\right)
\end{equation}
(see equation (\ref{eq:LzQ_Series_Third_Order_III})). The second factor in
equation (\ref{eq:factors-third-order-appendix}) is also obtained by working
in expansions of $\tilde{S}$, which proceeds by a simpler derivation (see
equation (\ref{eq:S2E2Q_Series_III})). The orbital inclination, $\iota$, is
then obtained by using equation (\ref{eq:iota-third-order-in-S}).

\pagebreak%

\end{document}